\documentclass[smallcondensed,final]{svjour3}  
\smartqed  
\usepackage{graphicx}
\usepackage{mathptmx}      

\journalname{Experimental Astronomy}

\begin{document}

\title{A real-time software backend for the GMRT}

\author{Jayanta Roy \and Yashwant Gupta \and Ue-Li Pen \and Jeffrey B. Peterson 
\and Sanjay Kudale \and Jitendra Kodilkar}

\institute{Jayanta Roy \at
              National Centre for Radio Astrophysics, TIFR, Pune University Campus, 
              Post Bag 3, Pune 411 007, India\\
              Tel.: +91-20-25719000\\
              Fax: +91-20-25692149 \\
              \email{jroy@ncra.tifr.res.in}           
           \and
           Yashwant Gupta \at
	      National Centre for Radio Astrophysics, TIFR, Pune University Campus, 
              Post Bag 3, Pune 411 007, India\\
              Tel.: +91-20-25719000\\
              Fax: +91-20-25692149 \\
              \email{ygupta@ncra.tifr.res.in}           
           \and
	   Ue-Li Pen \at
	      Canadian Institute of Theoretical Astrophysics, University of Toronto, 
              Canada\\
	      Tel.: +1-416-978-6477\\
	      Fax: +1-416-97803921\\
	      \email{pen@cita.utoronto.ca}
	   \and
	   Jeffrey B. Peterson \at
	      Carnegie Mellon University, Pittsburgh, USA\\
              Tel.: +1-412-268-2785\\
              Fax: +1-412-681-0648\\
              \email{jpb@cmu.edu}
	   \and	
	   Sanjay Kudale \at
              National Centre for Radio Astrophysics, TIFR, Pune University Campus, 
              Post Bag 3, Pune 411 007, India\\
              Tel.: +91-20-25719000\\
              Fax: +91-20-25692149 \\
              \email{ksanjay@ncra.tifr.res.in}
           \and 	
	   Jitendra Kodilkar \at
              National Centre for Radio Astrophysics, TIFR, Pune University Campus, 
              Post Bag 3, Pune 411 007, India\\
              Tel.: +91-20-25719000\\
              Fax: +91-20-25692149 \\
              \email{jitendra@ncra.tifr.res.in}           
}

\date{Received: date / Accepted: date}

\maketitle

\begin{abstract}
The new era of software signal processing has a large impact on radio astronomy instrumentation. 
Our design and implementation of a 32 antennae, 33 MHz, dual polarization, fully real-time software 
backend for the GMRT, using only off-the-shelf components, is an example of this.  
We have built a correlator and a beamformer, using PCI-based ADC cards and a Linux cluster 
of 48 nodes with dual gigabit inter-node connectivity for real-time data transfer requirements.  
The highly optimized compute pipeline uses cache efficient, multi-threaded parallel code, with the aid of 
vectorized processing. This backend allows flexibility in final time and frequency resolutions, and 
the ability to implement algorithms for radio frequency interference rejection. Our approach has allowed 
relatively rapid development of a fairly sophisticated and flexible backend receiver system for the GMRT, 
which will greatly enhance the productivity of the telescope. In this paper we describe some of the 
first lights using this software processing pipeline. We believe this is the first instance 
of such a real-time observatory backend for an intermediate sized array like the GMRT.  

\end{abstract}

\section{Introduction}
\label{intro}
Radio astronomy started with single dishes. The sensitivity and resolution of a single
dish radio telescope is limited by its physical aperture area. Due to the larger
wavelengths involved, the resolution of even the largest such radio telescope -- the
Arecibo dish, which is 300 m in diameter -- is poor compared to its optical counterparts.
To overcome this limitation, radio astronomers have evolved the technique of interferometry
which synthesizes large apertures using multiple single dishes. The first multi-element radio 
interferometer for astronomy was built by Sir Martin Ryle \cite{Ryle}. An interferometer measures the 
cross-correlation between the signals for a given pair of antennae. The instantaneous value of 
this cross-correlation (also called ``visibility'') estimates one Fourier component of the 
two dimensional brightness distribution in the sky plane.  The frequency of the measured Fourier 
component depends on the vector distance between the two antennae, projected onto a plane normal 
to the direction of the source (also called ``baseline"). For an array of $N$ antennae, at any given 
time, there will be $N*(N-1)/2$ such baselines and corresponding visibility measurements.  
Due to the rotation of the Earth, the baseline vectors change with time, allowing many more 
Fourier components to be measured.  By taking a 2-D Fourier Transform of the measured visibilities 
(after appropriate calibration), one can obtain the sky brightness distribution \cite{Thompson}, 
with a resolution that corresponds to an aperture size equal to the largest separation in the array.  
Some of the major aperture synthesis interferometer arrays in operation today are, the Very Large 
Array (VLA) of NRAO, the Westerbork Synthesis Radio Telescope (WSRT) of ASTRON, the Australia 
Telescope Compact Array (ATCA) of ATNF, the Giant Metrewave Radio Telescope (GMRT) of NCRA.

The GMRT consists of an array of 30 antennae, each of 45 m diameter, spread over a region of 25 km diameter, 
and operating at 5 different wave bands from 150 MHz to 1450 MHz \cite{Swarup}.  Amongst the 
multi-element earth rotation aperture synthesis telescopes operating at meter wavelengths, the GMRT
has the largest collecting area. 
The GMRT can also be configured in array mode, where it acts as a single dish by adding the signals from 
individual dishes \cite{Gupta}. This mode of operation is used for studying compact objects like pulsars, 
which are effectively point sources even for the largest interferometric baselines. The maximum instantaneous 
operating bandwidth at any frequency band is 33 MHz. Each antenna provides signals in two orthogonal 
polarizations, which are processed through a heterodyne receiver chain and brought to the central receiver 
building, where they are converted to baseband signals and fed to the digital backend consisting of 
correlator and pulsar receiver. The existing GMRT hardware backend uses  Application Specific Integrated 
Circuit based design approach, which is fairly efficient and optimized for its prime scientific goals, but 
has limited flexibility. It also suffers from some limitations due to quantization effects in the hardware 
pipeline.  The pulsar receiver, which uses Digital Signal Processing chips, is somewhat more flexible in 
its configuration and capabilities.  With the growing impact of the GMRT, there is a strong need felt by 
the user community for a flexible and more powerful backend which can support new regimes of observations. 
Some examples of these are : (i) spectral line observations requiring higher frequency resolution over larger 
bandwidths  (ii) pulsar searches and surveys for transient sources requiring higher time resolution across a 
wide field of view, including the capability of forming multiple phased array beams within the primary field 
of view.  Furthermore, like most low frequency radio telescopes, the GMRT is also affected by local radio 
frequency interference (RFI) -- both narrow spectral lines and broadband, impulsive signals. To produce data 
with high dynamic range and low noise, it is essential to detect and filter out such RFI signals, at various 
stages in the processing pipeline.  The existing GMRT hardware backend is not designed to address such issues. 
A software based backend can be an attractive option to overcome these limitations. 

A software backend is a real-time or off-line processing pipeline that runs on a supercomputer
or cluster of computers.  The rapid growth in general purpose computing power during the last
few years had made it possible to compete with the processing speeds of dedicated hardware. Besides
this, fast data transport links between computers are now possible due to the availability of
high speed gigabit networking. Storage media have also gone through a revolution in the last few
years, in terms of capacity and throughput. All these recent advancements have made it attractive
to attempt designs of software backends for radio astronomy.  This is aided by the fact
that the computing required for the processing of signals from a multi-element interferometric
array is very well suited for implementation on multi-processor computers. Use of commercial 
off-the-shelf (COTS) components and software processing blocks significantly reduces the several 
years of development time required for hardware backends.  It also makes the upgrades much easier, 
by replacing with faster computers without spending effort in rebuilding the processing blocks again. 
The algorithms implemented in software are much more flexible; new features can be added easily, and 
new regions of parameter space can be explored very conveniently. 

Design of correlators using off-the-shelf general purpose computers is not an entirely 21st
century approach. Very long baseline interferometric (VLBI) observations, where the distances
between the antennae are much greater than in conventional interferometers, were started with 
the recording of raw voltage samples from each antenna onto magnetic media. The recorded data 
from all the antennae in the VLBI network were correlated off-line using general purpose computers. 
The first such software correlator was implemented on IBM mainframes (system 360/50) at the NRAO by 
the NRAO-Cornell VLBI group around 1967 \cite{Bare}. The input data were from a Mark I tape 
recording system, albeit with a rather modest operating bandwidth of 360 kHz \cite{Kellermann}. 
Over the next few decades, the increase in complexity of the VLBI systems -- due to increase in 
bandwidths and number of antennae $-$ resulted in a switch towards specialized hardware 
correlators to process the data, e.g. Very Long Baseline Array (VLBA) of NRAO, Joint Institute 
for VLBI in Europe (JIVE), the Canadian S2 system.  This was the era where hardware/firmware 
options were found to provide realistic solutions \cite{Carlson}.  However, Moore's law was slowly 
catching up with these requirements, and by the late 1990s, the availability of computing power (including 
distributed parallel processing) and high speed, high density data storage capabilities renewed 
the interest of the radio astronomy community in software processing. In 1990, a gated cross-correlator 
for Mark II VLBI was devised by Petit et al. \cite{Petit}. The DiFX VLBI correlator \cite{Deller} 
designed by the Swinburne University of Technology, which extensively employs parallel processing 
techniques, is the most recent milestone in this regard. However, all these projects are mostly for 
off-line processing of recorded data.  Phillips et al. \cite{Phillips} demonstrated real-time 
e-VLBI for an array of four telescopes from the Australian Long Baseline Array, with 16 MHz of 
observing  bandwidth, using the DiFX correlator.  However, the work described here to develop a 
software based backend for the GMRT is probably the first implementation of a fully {\it real-time} 
software pipeline as a regular observatory backend for a medium sized array.
The upcoming BlueGene-L based central processing unit for the LOFAR telescope \cite{Romein} will be 
another such fully real-time backend. The IBM cell processor based e-VLBI correlator is also aimed for 
real-time data transport and software based processing \cite{Wagner}.

The recently developed GMRT software backend (GSB), built using mainly COTS components, is a
fully real-time backend that supports all the features of the existing hardware backend of the GMRT.
In addition, it provides substantially enhanced spatial and temporal resolutions.  Furthermore,
it supports a baseband recording mode where raw voltage signals from all the antennae can be recorded
to disk for specialized off-line processing to maximize the science return.  In this paper, we describe
our design and implementation of this 32 antennae, 33 MHz, dual polarization, fully real-time
software backend for the GMRT.

\section{Design of the GSB}
\label{sec:1}
The basic design requirements for the GSB are to support two main modes :
(i) a real-time correlator and beamformer for an array of 32 numbers of dual polarized
signals with a maximum bandwidth of 33 MHz, (ii) a baseband recorder for the above signals,
accompanied by off-line correlation and beamforming.  The details of the design concepts
and considerations for these modes of operations are described below.

\subsection{Correlator design considerations}
\label{sec:2}
At the GMRT, the dual polarized voltage signals from each antenna are processed through 
superheterodyne receivers and finally brought to a central location for further processing. The
intermediate frequency signals from each antenna are then down-converted to baseband signals 
and fed to the digital signal processing backend. Within the backend, these signals need to be 
Nyquist sampled, and given to a spectral correlator that can estimate the visibility products
for each baseline, for multiple spectral channels. The spectral correlator can be 
of either FX or XF type, depending on whether the cross-multiplication is done after or before 
the frequency analysis \cite {Thompson}. For the GSB, a FX correlator approach is utilized, 
where the input signals first get channelized using a Fast Fourier Transform (FFT) implementation, 
before being cross-multiplied and accumulated. In order to preserve good dynamic 
range in presence of RFI, a minimum of 4 bit sampling is required.  At the lower frequencies of
the GMRT, where the RFI is more severe, the usable bandwidths are smaller, and 8 bit sampling is
required.  The delay correction, $\tau_{tot}$, required to time align the sampled data 
streams from the different antennae has two main contributions : (i) the geometrical path delay 
of the wavefront reaching a given antenna (with respect to a reference antenna), $\tau_{g}$, 
which depends upon the direction of the source ($\hat{S_{0}}$) and the baseline vector 
($\vec{D_{\lambda}}$) between the two antennae as  
(see pages 68-73 of \cite {Thompson} for detailed descriptions and figures)
\begin{equation}
\tau_{g} = \frac{1}{c} (\vec{D_{\lambda}} . \hat{S_{0}})
\end{equation}

and (ii) the fixed delay, $\tau_{fix}$, which arises due to the fixed path length
differences from the antenna to the backend (mostly cable lengths plus other
instrumental delays). A GMRT specific delay model \cite{Jayaram} is used to calculate this 
total antenna based delay and its rate of change (${\dot{\tau}}_{tot}$). For the GMRT, the GSB needs
to provide a maximum tunable delay $\tau_{tot}$ of 128 $\mu$s, with an accuracy of 2 ns \cite{Jayaram}
to match the performance of the hardware correlator. 

In a digital backend, the delay correction implementation is decomposed into two parts. The first 
part is an integer delay ($\tau_{I}$) correction, which is applied in integer multiples of the 
sampling clock, and results in the data streams being aligned to within $\pm$ 1 sample. It is usually 
implemented by block shifting of the sampled time series data streams. The second part is a fractional 
sample time correction (FSTC), which takes care of the residual delay of less than one sample interval 
($\tau_{frac}$). In a FX correlator, it is usually implemented in the Fourier domain by multiplying 
the spectral channel data with a phase gradient. 

In addition to the above mentioned delay corrections, there is an additional phase correction of the 
signal from each antenna, called fringe correction (or fringe de-rotation), that needs to be carried out. 
This is a broad-band phase correction which arises from the fact that the $\tau_{tot}$ is corrected at 
a baseband frequency, $\nu_{BB}$, instead of the sky frequency, $\nu_{RF}$.  The total phase 
correction as a function of frequency and time is
\begin{equation}
\phi(\nu, t) = 2\pi([\nu_{BB}+\nu_{i}]\tau_{frac} \pm [\nu_{RF} - \nu_{BB}]\tau_{tot})
\end{equation}
where $\nu_{i}$ is the baseband frequency of the $i^{th}$ spectral channel, $\nu_{RF}$ is the sky frequency 
corresponding to the lower edge of the band, and $\nu_{BB}$ is the baseband frequency at the lower edge. 
The difference between these two is the local oscillator frequency, $\nu_{LO}$, used for the frequency
translation.  The ``$\pm$'' sign signifies the passband convention  (``$+$'' for $\nu_{RF}$ $>=$ $\nu_{LO}$ 
and ``$-$'' for $\nu_{RF}$ $<$ $\nu_{LO}$).  
Fringe correction can be implemented either before or after the FFT.  In pre-FFT fringe correction, the
time domain voltage signal is multiplied with a phase correction factor, whereas in post-FFT fringe 
case, the correction is carried out by multiplying the spectral channel data of each time sample 
with a complex phase factor.   Post-FFT fringe correction (combined with FSTC correction) is 
computationally advantageous and works well in cases of low fringe rates and relatively small number 
of spectral channels, such that the change in fringe phase over one FFT cycle is negligible. 
For a low frequency, moderate array size telescope like the GMRT, the maximum fringe rate is $\pm$ 5 Hz.  
For FFT length of 1024 points, this results in a worst case decorrelation of 0.18$\%$ due to the
variation of the fringe phase during one FFT cycle, which is quite acceptable.  Hence, for the GSB, 
post-FFT fringe correction is sufficient.

A FX correlator computes the instantaneous cross power spectrum between every pair of antennae, 
by cross multiplying the FFT data from the corresponding frequency channels of the selected antennae. 
These products are integrated for the desired time scale to obtain the time sequences for the complex 
visibilities for different spectral channels, for all baselines. This multiply-accumulate (MAC) operation 
is repeated for data from each of the polarizations, e.g. right-circular polarization (R) and 
left-circular polarization (L). Beside these total intensity products, the GSB, when in full polar mode, 
also needs to provide all the four products ($RR^{*}$, $RL^{*}$, $LR^{*}$, $LL^{*}$) required to reconstruct 
the Stokes parameters.  For the GSB, the default value of the desired time resolution for the final visibility 
data is 2 s.  In some special cases, the GSB is required to output the visibilities at a faster rate.

For a correlator, the computational costs are as follows.  For the FFT block, for $N$-point transforms
of signals from $N_{a}$ antennae with $N_{p}$ polarizations over a total bandwidth of $\Delta\nu$ (in Hz), 
the computational load is $2N_{a} N_{p}\Delta\nu$$log_{2}N$ complex operations per second (Cops).  The 
computational cost for the post-FFT fringe and FSTC corrections is a further $N_{a} N_{p}\Delta\nu~$ Cops. 
For the MAC operations, the requirement is $\Delta\nu N_{s} N_{a} (N_{a}+1)~$ Cops, where $N_{s}$ 
is the number of polar products per baseline.  Using a conversion factor of 4 for Cops to real floating point
operations (flops), this works out to a compute requirement of 181 Gflops for 2048 point spectral decomposition, 
8.25 Gflops for the phase correction, and 280 Gflops for the MAC operations, in total intensity mode, for the 
GMRT specifications. 

\subsection{Array beamformer design considerations}
\label{sec:3}
A beamformer is effectively a highly sensitive single dish built from an array of telescopes. For the GMRT 
array, the beamformer specification is to provide two modes of operations : (i) an incoherent array mode, 
where the voltage samples from the selected antennae are added after converting to intensities, (ii) a 
coherent or phased array mode, where the voltage signals from the selected antennae are first added, and 
then converted to intensity samples \cite{Gupta}. Both these modes require antenna based gain offsets and 
time delays to be corrected prior to the addition of the signals. Furthermore, for the coherent beamformer 
mode, it is also required to calibrate out antenna based phase offsets \cite{Sirothia}, both for broadband 
and narrowband spectral variations.  The incoherent array beam has the same field-of-view as the primary 
beam of a single antenna, but with an enhanced sensitivity of $\sqrt{N_{a}}$ times that of a single antenna, 
for an array of $N_{a}$ antennae.  However, the coherent array beam is much narrower than that of a single 
antenna, being similar to the synthesized beam obtained from the array of $N_{a}$ antennae, and has a 
sensitivity improvement of $N_{a}$ times than that of a single antenna. Thus, the incoherent array mode 
is more useful for rapidly covering large areas of the sky, as would be needed in large scale pulsar surveys, 
whereas the coherent array mode is ideal for studies of known pulsars and for high-sensitivity searches for 
pulsars in compact targets such as globular clusters and supernova remnants.

Both the incoherent and coherent beamformers produce total intensity signals as the final outputs. In 
addition to this, the coherent beamformer is also required to give all the voltage cross products needed 
to derive the Stokes parameters of the incoming radiation.  For pulsar observations, incoherent and 
coherent array outputs need to be corrected for the effect of dispersion in the interstellar medium (ISM). 
In most cases, the dedispersion process employs incoherent dedispersion, which corrects for the pulse 
smearing across the frequency band, but is limited by the dispersion delay within the spectral channels. 
For exact correction, the Nyquist sampled pre-detected voltage signals from the phased array output can 
be put through the coherent dedispersion technique which preserves the full time resolution of the data 
\cite{Hankins}. 

For an array of $N_{a}$ antennae each with a bandwidth of $\Delta\nu$ (in Hz) and $N_{p}$ 
polarizations, the incoherent beamformer cost is $N_{a} N_{p} \Delta\nu$ (multiplication) Cops 
$+$  $N_{a} N_{p} \Delta\nu$ (addition) Cops.  For the GMRT, this translates to 12.5 Gflops.  
Similarly for the coherent beamformer, for $N_{b}$ beams (covering different parts of the 
primary beam), each with $N_{s}$ polar products, the cost is 
$(N_{a} + N_{s}) N_{p} N_{b} \Delta\nu$ (addition) Cops $+$ $N_{p} N_{b} N_{s} \Delta\nu$ (multiplication) Cops. 
For a single beam in total intensity mode, this is equal to 4.5 Gflops. The beamforming costs are thus 
negligible compared to the correlation costs.

\subsection{Overview of the GSB architecture}
\label{sec:4} 
The GSB implements a high performance computing platform using off-the-shelf commodity machines.
A highly optimized software pipeline has been developed to achieve the design goals of the GSB, 
with a minimum of hardware investment. The details of the hardware and the software architectures 
of the GSB are described below.

\subsubsection{The GSB cluster}
\label{sec:5}
\begin{figure}
\begin{center}
  \includegraphics[angle=0,width=1.0\textwidth]{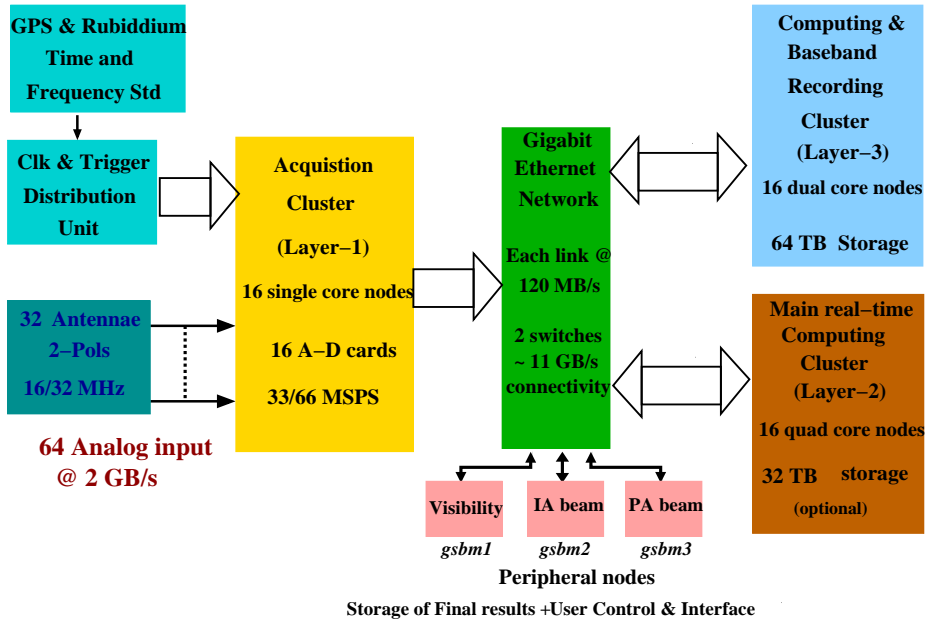}
\caption{The basic block diagram of the GMRT software backend. This shows the 3 kinds of nodes (labeled
as Layer-1, Layer-2 and Layer-3) and their inter-connectivity via a dual gigabit ethernet network, to
which the peripheral machines are also attached.  The input baseband signals to the
backend, and trigger and clock distribution arrangements, are illustrated by blocks on the left.}
\label{fig:1}
\end{center}
\end{figure}

A schematic view of the configuration of the GSB cluster is shown in Fig. \ref{fig:1}.  The basic 
design uses a Linux cluster of 48 Intel Xeon nodes, which are interconnected via two independent 
48-port commercial gigabit switches with sustained switching bandwidth of 11 GB/s. The nodes are 
segregated into three kinds (called layers), in terms of compute capabilities and usage.  Layer-1 
consists of 16 acquisition nodes which are Xeon 2.4 GHz single core dual processors, each having 
a 64 bit, peripheral computer interface (PCI) based four channel analog to digital converter (ADC) 
card.  This card can acquire data from 4 analog inputs, operating at either 4-bit, 33 MHz or 8-bit, 
16 MHz bandwidth mode.  This results in a sustained data throughput rate of 132 MB/s on the PCI bus 
of each acquisition node. The total input data rate to the acquisition cluster is thus 2 GB/s. The 
16 acquisition nodes, each handling 4 analog inputs, provide a 64 channel capability to the GSB, 
which takes care of the 30 antennae, dual polarization requirement of the GMRT, while still leaving 
scope for 4 additional test signals to be connected.  Each ADC board is equipped with on-board 
phase-lock-loop and trigger logic to synchronize the ADC sampling clocks and start of acquisition 
across all the 64 inputs.  The clock synchronization reference signal is derived from the 
observatory's frequency standard, which is a Rubidium atomic clock, and the trigger synchronization
signal is derived from the time standard, which is a GPS receiver, and these are distributed to 
each of the 16 boards, as shown in Fig. \ref{fig:1}.  There are programmable variable gain amplifiers 
at the input of each ADC, which provide a facility to digitally control the gain of each antenna's 
baseband signal.

Layer-2 consists of 16 numbers of Xeon 2.3 GHz quad core dual processor nodes. These nodes
handle the bulk of the computational load of the main processing pipeline.  Layer-3 consists 
of 16 numbers of Xeon 3 GHz dual core dual processor nodes, each with 4 TB of SATA disk storage. 
These nodes primarily work as a recording cluster in the raw dump mode, but also take part in 
the computations, for modes with higher compute requirement than can be handled by layer-2 alone
(e.g. 33 MHz full polar mode of operation). Each of the 48 nodes in the cluster has dual gigabit 
on-board ethernet links for establishing inter-node connectivity.

In addition to the 48 acquisition and compute nodes which form the core of the cluster, there are
a few peripheral nodes attached to the system.  The main gateway to the GSB is a manager
node (called {\it gsbm1} in Fig. \ref{fig:1}), which provides the primary interface to the outside 
world.  It provides the control and configuration information from the central control software system 
of the GMRT, which includes details of the antenna connectivity to the acquisition nodes, current frequency 
and source settings of the antennae, antenna specific gain and phase updates, command signals to start the
data recording for a new scan, and other related operations.  This node also receives the final
visibility results from the GSB cluster and passes them to a machine that does the long term accumulation
and records the data files on disk. These files are finally converted to standard FITS files, which can 
be loaded into the AIPS \footnote{see http://www.aoc.nrao.edu/aips} data analysis package.   
There are two other machines (called {\it gsbm2} and {\it gsbm3} in Fig. \ref{fig:1}) that are attached 
to the cluster to receive the incoherent array and phased array beam data. These nodes record the beam 
data to local disks after some preprocessing, as required. 

\subsubsection{Software architecture of the GSB}
\label{sec:6}
The GSB code is a parallel pipeline running on the 48 cluster nodes and the peripheral machines, 
described in Sec \ref{sec:5}.  Since the different layers of nodes are performing different but 
inter-related jobs, proper synchronization between all of them is required. Efficient transmission of 
large volumes of data between them is also needed.  These are achieved by using Message Passing Interface 
(MPI)\footnote{see http://www.open-mpi.org/} as the main tool for communication and synchronization 
between the nodes.  Special care is taken to ensure that all the nodes reach well defined 
synchronization barriers after processing a specific block of data. In addition, OpenMP\footnote{see 
http://openmp.org/} based multi-threading techniques are used on the computing nodes in order to optimize 
and balance the different computing tasks required to be performed on the data. Further, Intel 
IPP\footnote{see http://software.intel.com/en-us/intel-ipp/} routines and vector programming are 
exploited to get the best performance from the compute nodes.

\begin{figure}
\begin{center}
  \includegraphics[angle=0,width=1.0\textwidth]{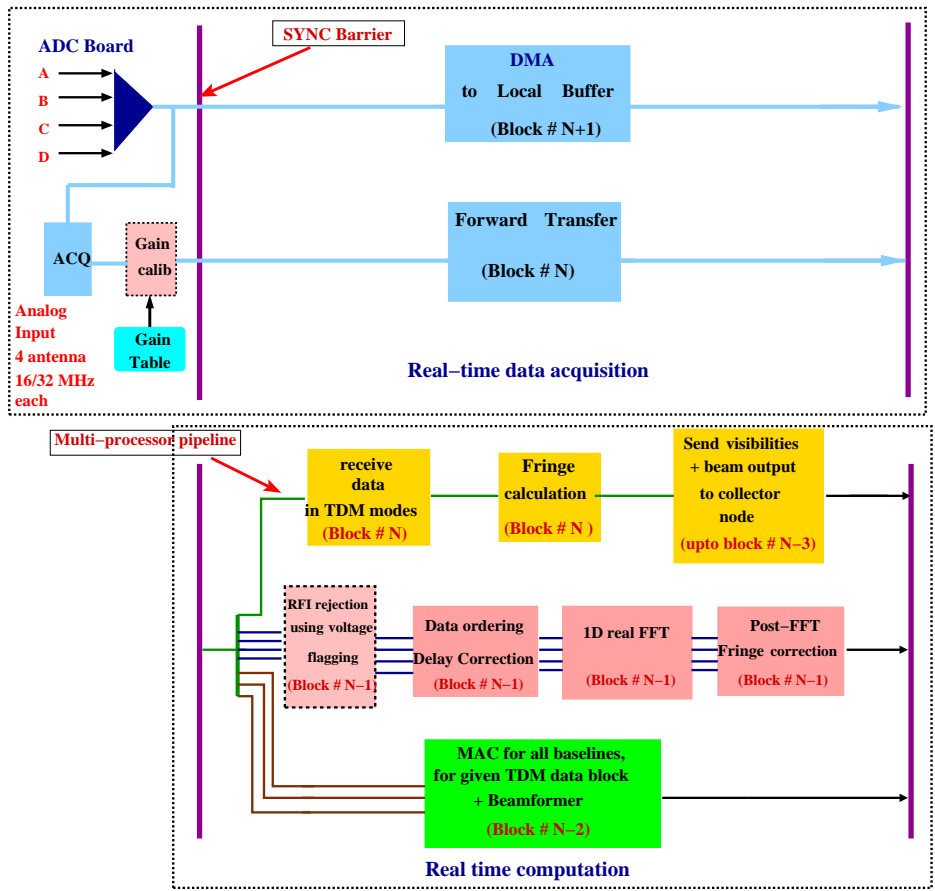}
\caption{The software tool flow of the real-time GSB code : the upper panel describes the real-time data
acquisition part, illustrated for a single acquisition node; the lower panel describes the real-time
computation part, illustrated for a single compute node.  The parallel sections are bounded by synchronization
barriers, indicated by the vertical lines.  The different blocks between these barriers signify different
compute tasks.  The number of lines joining to a block is the number of threads assigned for that compute task.}
\label{fig:2}
\end{center}
\end{figure}

Fig. \ref{fig:2} shows the software flow of the GSB code. The upper panel shows the software flow 
for the real-time data acquisition part running on the layer-1 nodes, while the lower panel shows 
the software flow for the real-time computing part running on the layer-2 / layer-3 nodes.  The
bulk of the code runs in a continuous loop in parallel sections (delineated by the vertical lines
in Fig. \ref{fig:2} that mark MPI synchronization barrier points) that execute on both kinds of nodes,
except for a small part that runs sequentially, once in the beginning, on the layer-1 nodes (on 
the left of the synchronization barrier in Fig. \ref{fig:2}).  This sequential part of the code 
performs the initial set-up of the acquisition cards (depending on the mode of operation selected 
by the user) and arms the cards to trigger the data acquisition at the next GPS pulse. It also allows 
for setting the gain for each input signal using a gain calibration table, which can be adjusted 
to optimally exercise the 8-bit samplers.  After that, the code enters into the parallel sections 
that execute between successive MPI barrier points, on all the nodes.  On the acquisition nodes, the 
main tasks for this part of the code are to poll for new buffer of data from the ADC board and to 
transfer the buffer to the compute nodes. The data are transferred from the on-board memory of the 
ADC card to the local memory of the acquisition nodes via interrupt driven direct memory access. 
The data rate achieved for these transfers is 150 MB/s.  The size of each buffer received by the 
acquisition nodes is 32 MB, consisting of 8 MB from each of 4 antennae input signals.  This acquisition 
block corresponds to a time slice of 251 ms, for both the 16 MHz, 8 bit mode and the 33 MHz, 4 bit mode
of the GSB. The timestamp for each data block is derived from the local GSB time-server which is tied 
to the observatory time server through Network Time Protocol (NTP). The acquisition nodes use a double 
buffer ping-pong scheme : while data for one block are being acquired in one buffer, the data from the 
previous acquisition block (sitting in the other buffer), are transferred to the compute nodes.  The 
data for each block from the 16 acquisition nodes is transferred to the 16 compute nodes in a 
time-division-multiplex (TDM) mode wherein each compute node receives one time slice (of 1/16$^{th}$  
of a data block) from all the nodes, containing data of 15 ms duration.  This TDM data sharing scheme
ensures that all computations required for obtaining the visibilities for all baselines as well as the 
beam outputs, for a given time slice, can all be carried out on a single compute node.  
For the modes of the GSB where extra computing load is needed, requiring the layer-3 nodes also to 
participate, the above concept is extended to generate 32 time slices of the original buffer of data, 
which are processed by the 32 compute nodes of layer-2 and layer-3. The length of the time slices 
going to layer-2 and layer-3 nodes are adjusted to cater to the different compute capabilities of the 
nodes in these two layers. 

On each compute node, we run parallel pipelines using an OpenMP based multi-threaded environment. 
This is done in order to optimally utilize the multiple cores of the compute nodes. The number of threads
is kept equal to the number of cores. For some tasks, such as receiving the data from the layer-1 nodes
over the network, a single thread resource is sufficient to carry out the job.  In other cases, such as
the FFT and MAC operations, 3 to 4 threads are found to balance the load most optimally. In most
of these cases, devoting all 8 threads to the same task was found to be sub-optimal. Hence, the 
software flow on each compute node has 3 multi-threaded parallel pipelines (as shown in the lower panel 
of the Fig. \ref{fig:2}), as opposed to a model where all the jobs are done sequentially by a single 
multi-threaded pipeline.  To make this work, the parallel sections operate on different blocks of data
(labeled as block \#s N, N-1, N-2 and N-3 in Fig. \ref{fig:2}), corresponding to different time slices.  
This requires the data blocks to be buffered at the end of each of the main stages of operations : after 
network transfer, after FFT, and after MAC and beamforming operations.  Each of this is a double buffer 
that is used in a ping-pong manner. 

The details of the 3 parallel sections running on the compute nodes are as follows. The first is a 
single thread section (the top pipeline shown in the bottom half of Fig. \ref{fig:2}) that handles 
handshaking with the layer-1 nodes and the network transfer in TDM mode, calculation of all delay and 
fringe parameters (as described in Sect. \ref{sec:2}) for the current time slice (labeled as block \# $N$) 
being processed. In addition, this pipeline also transfers the reduced results 
(e.g. visibility and beam data) from the ($N-3$)$^{th}$ block to the corresponding peripheral nodes. 
The second pipeline is a section with 4 threads (middle computation pipeline shown in Fig. \ref{fig:2}) 
that performs data reordering (demultiplexing the 4 antennae data streams from each acquisition node 
and unpacking 4 bit data samples into 8 bits for the 33 MHz, 4 bit mode), integer delay correction, FFT 
and post-FFT fringe and FSTC correction $-$  all of these on the ($N-1$)$^{th}$ block.  The integer delay 
correction is implemented by offsetting the memory read pointer of the array when it is read in by 
the FFT routine. In order to have seamless delay correction of data samples between different time slices 
on different nodes, there needs to be some overlap of data samples from successive time slice blocks $-$ 
this is ensured during the network transfer of data from the acquisition nodes to the compute nodes. 
The number of time samples in the overlap section is predicated by the largest delay that needs to be 
compensated, and it set to cover $\sim$ 240 $\mu$s, which is more than the maximum requirement for the 
GMRT array, specified in Sect. \ref{sec:2}.  The post-FFT fringe and FSTC corrections are carried out 
by using precomputed sine and cosine values at discrete phase steps, stored as a look-up table.  
The table has 1440 steps, allowing a delay correction accuracy of 0.042 ns, more than the requirement 
specified in Sect. \ref{sec:2}. This pipeline also has an optional component (shown by the 
dotted box in Fig. \ref{fig:2}) which implements some simple RFI removal algorithms, described in
detail in Sect. \ref{sec:18}.  

The third pipeline (lower computation pipeline shown in Fig. \ref{fig:2}) is a section with 3 threads 
that carries out the MAC and beamforming operations, working on the ($N-2$)$^{th}$ data block. Depending 
on the operating mode of the GSB that is selected, the MAC section produces either self-polar or 
full-polar visibilities for all the baselines, integrated over the time slice duration of the block, 
which is 15 ms.  The beamformer produces one incoherent and one coherent array beam at the raw FFT 
time resolution, covering the duration of the time slice.  Depending on the selected mode, the coherent 
array beam data can be the raw voltages for each polarization, or the self-polar intensities for each 
polarization, or the full-polar intensities. The visibility data need to be further integrated over the 
different time slices from each node. To overcome the many-to-single network congestion that would occur 
if the visibilities were sent from each compute node directly to the peripheral manager node, we have 
employed a tree based reduction technique, where nearby compute nodes form groups to integrate data within 
themselves, during successive iterations of the parallel section. Finally the local group heads send 
the reduced volume of data to the manager. For the beamformer, in order to preserve the high time 
resolution, the data from successive time slices can not be integrated.  Instead, the full resolution 
data are sent out sequentially from the individual nodes of the cluster to the receiving peripheral nodes, 
one each for the incoherent and coherent array beams.  As we use different peripheral nodes for
different beams, this helps us to separate out the network transmission paths for different beams. 
The maximum output rate for the integrated visibility data is 4 MB/s and for the beamformer data it 
is 128 MB/s for each of the beams, if only total intensity samples are sent.  For the pre-detected voltage 
data in the phased array beam, the output rate is also maintained at 128 MB/s by reducing the number of bits 
per sample.  In the default conditions, the GSB produces 512 spectral visibility products for all baselines 
at a time resolution of 2 s, and single incoherent and coherent beams with a time resolution of 30 $\mu$s, 
for both the 16 and 33 MHz bandwidth modes of operation.

The total theoretical {\it scalar} compute power of the 16 layer-2 nodes is 295 Gflops, which is less than 
the total requirement, even for the basic modes of operations, which is $\sim$ 490 Gflops (Sect. \ref{sec:2} 
and \ref{sec:3}).  Further, since these are highly data intensive operations, there are significant 
overheads due to frequent and large volume data input-output (I/O) operations to and from the memory, 
which increase the disparity between available and required capabilities.  Use of the 16 layer-3 nodes 
for computing can partially takes care of the problem. However, their main role is to enable the raw data 
recording (and read-back) mode of the GSB.  In order to support the 24$\times$7 observatory backend mode 
of the GSB, where a raw data recording run can be immediately followed by a real-time correlation run 
(during which, the layer-3 nodes could be busy in analysis or re-transmission of the earlier recorded data), 
it became important to fit the real-time computing completely on the layer-2 nodes.  This required significant 
amount of optimization of the code, using techniques such as vectorized processing to utilize the full 
{\it vector} compute power of the cluster, which can be a factor of 8 more compared to its scalar capabilities, 
for the 16-bit arithmetic-logic-units (ALUs). In this context, use of fixed point processing over floating point 
processing provides a significant enhancement in computing power.  Optimizing the memory performance
by reducing the memory I/O overheads to a minimum, is also an important part of the code optimization.
We discuss all these issues in detail in the Sections \ref{sec:11} and \ref{sec:12}.

\subsection{The GSB baseband recorder}
\label{sec:7}
\begin{figure}
\begin{center}
  \includegraphics[angle=0,width=1.0\textwidth]{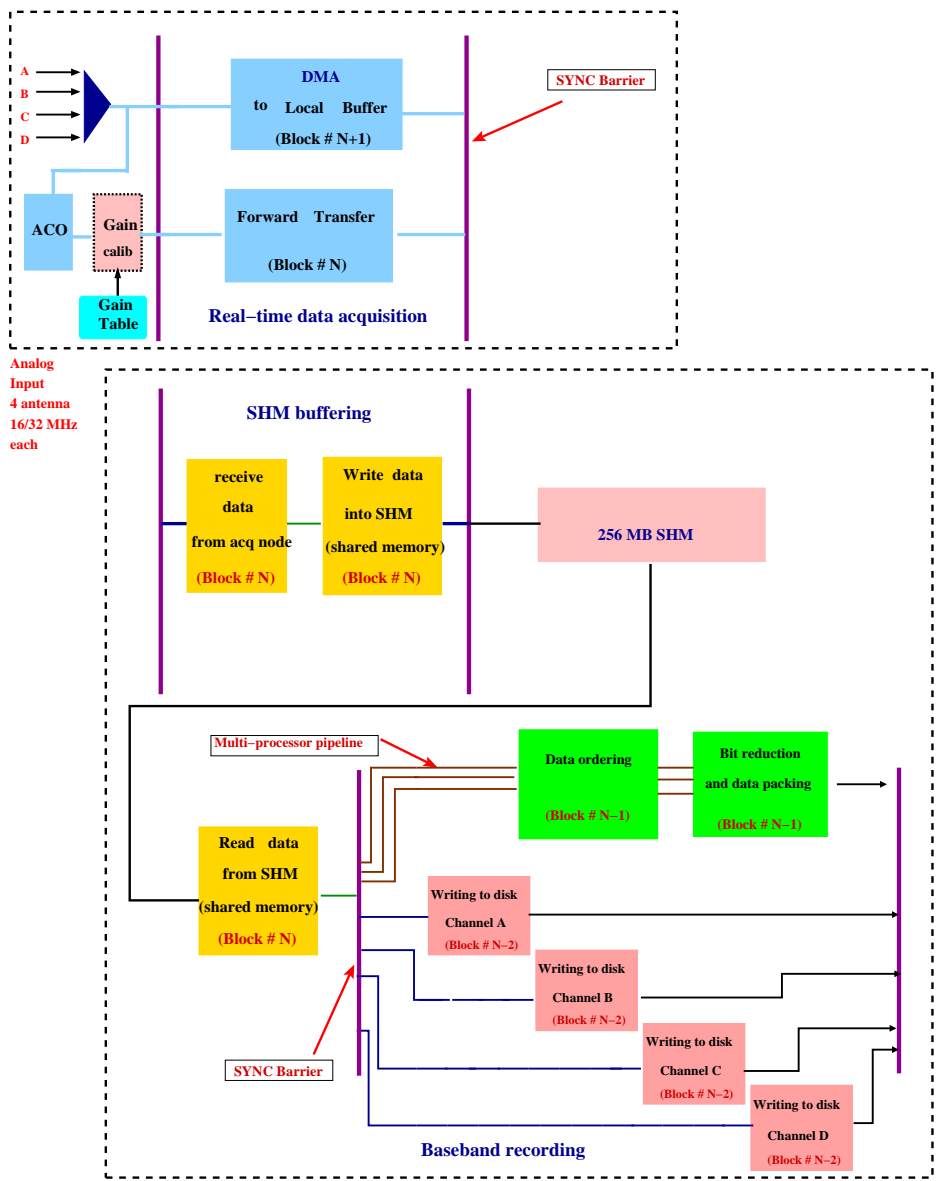}
\caption{The baseband recording software tool flow : the upper panel describes the real-time data
acquisition part, illustrated for a single acquisition node; the lower panel describes the baseband recording
part, illustrated for a single recording node. The parallel sections are bounded by synchronization
barriers, indicated by the vertical lines. The different blocks between these barriers signify different
tasks.  The number of lines joining to a block is the number of threads assigned for that task. There is
a 256 MB shared memory ring buffer, which is used to connect the acquisition and recording processes.}
\label{fig:3}
\end{center}
\end{figure}

In the baseband recording mode, the GSB cluster supports the streaming of raw voltage samples (at Nyquist
rate) from the antennae into an array of storage disks. The ADC samples from a given acquisition node 
travel directly to its recording counterpart in layer-3, through the dual gigabit ethernet connections. 
The recording cluster of 16 nodes, each with 4 TB of storage, provides a total storage capacity of 64 TB, 
which can support recording for 18 hrs of observations.  Similar to the compute pipeline for the real-time 
mode, the pipeline for the baseband recording mode is also based on an OpenMP, multi-threaded environment, 
bounded by software synchronization barriers.  Fig. \ref{fig:3} shows the software flow for the GSB 
baseband recorder.  The acquisition part of the code (shown in the upper panel of Fig. \ref{fig:3}) which
runs on the layer-1 nodes, is very similar to the real-time acquisition code (as described in the 
Sect. \ref{sec:6}), except for the scheme used for transfer of data over the network.
Instead of time slicing the data buffer in TDM mode, layer-1 nodes initiate pair-wise transfers, where 
each acquisition node sends the full block of data of 32 MB to its recording counterpart. 


On a recording node (shown in lower panel of the Fig. \ref{fig:3}), there are 2 different processes running, 
that are connected by a common shared memory ring buffer.  There is a single thread sequential pipeline to 
receive the current data block (\# $N$) from the corresponding acquisition node and to write the data into the 
ring buffer having a depth of 8 data blocks (i.e. total size of 256 MB). This process is bounded by a synchronization 
barrier, common with the acquisition nodes.  Data buffering using a shared memory ring buffer is 
a very useful technique for smoothing out the occasional glitches in disk I/O performance, and helps 
significantly in avoiding loss of data, and a consequent loss of synchronization with other data
streams.  The concurrent process for reading the shared memory uses a single thread to read one
block of data at a time, and then branches into 5 parallel sections (bottom half of Fig. \ref{fig:3}).
The upper pipeline uses 3 threads to perform data reordering as described in Sect. \ref{sec:6}, 
reduction of the data samples to 4-bit or 2-bit (for the 16 MHz and 33 MHz modes, respectively),
and packing of 2 or 4 such samples to construct a byte -- all of this on the previous block (\# N-1) 
of data .  The lower 4 pipelines are designed to perform the disk writing, where each single thread 
takes care of data for one antenna.  There are 4 numbers of 1 TB disks connected on each node, 
and data from each antenna streams into a separate disk.  Each of the disk I/Os performs at a sustained 
rate of 80 MB/s with the aid of a XFS filesystems \footnote{see http://xfs.org/index.php/Main\_Page}.  
Each recorded data buffer is accompanied with a timestamp derived from the NTP server, in addition to 
the synchronization of the main start of acquisition with a GPS minute edge.

The recorded data samples along with timestamps are processed off-line. The GSB cluster (layer-2 and/or 
layer-3) can be configured to play the role of the off-line analysis cluster.  We have also successfully 
ported the GSB off-line analysis pipeline on the main compute facility at NCRA $-$ a 72 node, 230 Gflops 
(theoretical scalar peak), Itanium cluster with Infiniband (IB) inter-node connectivity.  This cluster
is linked to the GSB with a dedicated 8 Gbps fibre link, which allows for close to real-time transfer
rates for the baseband recorded data.  The data is loaded into a network attached storage of $\sim$ 32 TB,
connected with the Itanium cluster.   The scalar compute power of the Itanium cluster is comparable to
that of layer-3 of the GSB, and it benefits similarly from vector optimization and fixed-point 
implementation techniques.   However, in overall performance, the Itanium cluster performs 
significantly slower than the layer-3 cluster of the GSB, as it is fed with data from a single 
storage unit, instead of the distributed storage of the GSB. We have 
also ported the GSB code on the Green Machine supercomputer at the Center for Astrophysics and 
Supercomputing of the Swinburne University of Technology.  The off-line pipeline running on any 
of these clusters is a floating point version of the real-time code, and employs all possible 
instances of the code optimization (as described in detail in the next section).

\section{Optimization techniques}
\label{sec:8}
High performance computing involves breaking large amount of data into smaller blocks and then 
performing calculations in parallel on those data blocks. Once these calculations are completed, 
the results are funneled to other processes that use them as input. The data passing between 
processes is handled by MPI. In order to turn Moore's law into actual performance gain, 
modern multiprocessor architectures include performance boosting features like multi-level 
caches, data prefetching, multiple execution units and special instruction sets 
for compute-intensive operations. 

\subsection{Efficient compute library}
\label{sec:9}
\begin{figure}
\begin{center}       
\includegraphics[angle=270,width=0.6\textwidth]{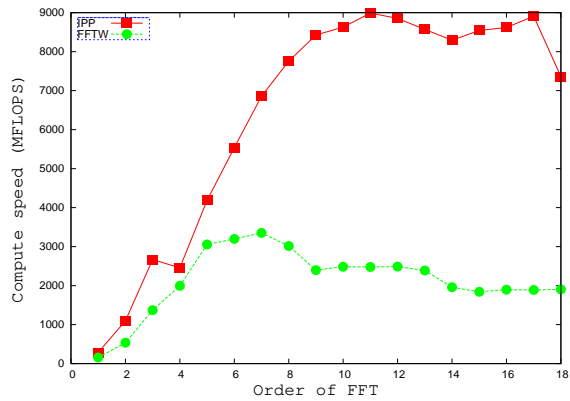}
\caption{Comparison of performance of the single-precision 1D FFT routine of the Intel IPP library 
with the FFTW routine. These benchmarks are obtained by running the routines on a single processor 
of one compute node of the GSB cluster.}
\label{fig:4}
\end{center}       
\end{figure}

The Intel processor's multi-core resources can be used optimally with the aid of optimized libraries 
such as the Intel Integrated Performance Primitives (IPP) library.  This has helped us significantly
to improve the performance of our code, compared to the use of other general purpose signal 
processing libraries (e.g. FFTW \footnote{see http://www.fftw.org/}). 
Our benchmark results show that the IPP based 1-D single precision real-to-complex FFT is more 
than a factor of 3 times faster than the FFTW, for transform lengths of interest to us 
(Fig. \ref{fig:4}).  Our real-time processing pipeline uses the IPP based 16-bit fixed point 
1-D real FFT.  This is found to achieve the same compute throughput as the IPP 32-bit floating 
point FFT, as it uses floating point arithmetic for the internal butterfly stages.  However, the
memory I/O throughput is reduced by a factor of two when using the fixed point version.

\subsection{Network optimization}
\label{sec:10}
The sustained real-time performance of the cluster requires stable, high speed data sharing between the
three layers of nodes. The network performance of the cluster has been optimized and runs at a sustained
rate of 240+ MB/s between any pair of nodes, using the dual gigabit connectivity offered by the two
switches.  This throughput comfortably takes care of all the real-time inter-node data transfer requirements $-$
both for forward transfer of data from the acquisition to the compute nodes and for transfer of results back from
the compute nodes to the peripheral machines.  The following three main aspects were found to be of importance
in the network optimization.

\begin{figure}
\begin{center}
  \includegraphics[angle=270,width=0.6\textwidth]{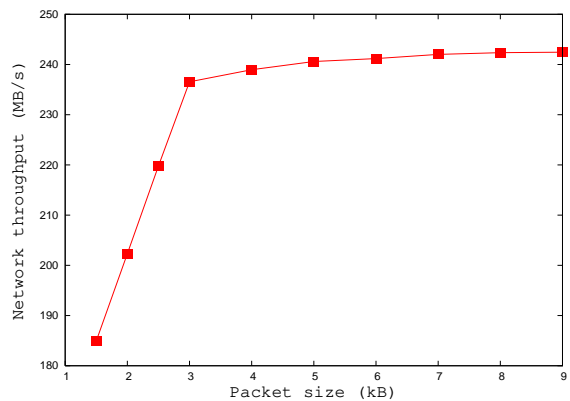}
\caption{Illustration of the dependence of network bandwidth on ethernet packet size. Optimized network
bandwidth is obtained for packet size of 8 KB.}
\label{fig:5}
\end{center}
\end{figure}

Firstly, in a multi-processing environment, dedicating the network interrupt request to a given processor
(called IRQ affinity) improves the resource sharing across various parallel processes. In our code, a
single OpenMP thread handles all the network transfers, and at the hardware level all the network
interrupts are localized to a single processor. Secondly, since the host processor is interrupted for 
every network packet arrived, a reduction of the rate of interrupts helps to improve the transfer efficiency. 
This is controlled by adjusting the packet size or Maximum Transmission Unit (MTU). By the use of 
``jumbo frames'', where the MTU size was increased from the default value of 1.5 KB to upto 8 KB, the 
overheads were reduced significantly, achieving as much as 30$\%$ increase in network bandwidth 
(shown in Fig. \ref{fig:5}). All the network transfer in the final GSB code runs with a 8 KB packet size.
Thirdly, optimal network performance was also found to depend on the mode of transfer and the MPI
communication protocol used. We found that the MPI many-to-many communication routine written using 
simultaneous pair-wise point-to-point nonblocking send-receive ({\it MPI$\_$Isend} and {\it MPI$\_$Irecv}) 
provides 5$\%$ more throughput than the available collective communication calls like {\it MPI$\_$Alltoall}.  
Further, we found that uni-directional transfer rates are 6$\%$ more than the bi-directional transfer rates, 
and most importantly, the sustained network performance is much more stable and reliable.  This was one of 
the main motivations for going for a separate layer of simple, inexpensive acquisition nodes, as compared 
to a model where the acquisition and computing is on the same set of nodes, as the former results in all 
the data transfers being unidirectional.

\subsection{Code optimization : load balancing and cache blocking}
\label{sec:11}
In order to fully utilize the power of the multi-core processors, proper load balancing using multiple
threads is a very important factor.  Care has to be taken during optimization to reduce the effect of 
thread synchronization delays.  The multi-threaded parallel code of the GSB uses the following programming 
model :

1. Domain decomposition or thread-level parallelism : The compute intensive tasks (e.g data reordering, FFT,
   fringe and FSTC corrections) in a given pipeline use the data divided into small subsets, where each thread
   can process smaller pieces of data independently.

2. Functional decomposition or task-level parallelism : The tasks with different functionalities are distributed
   on different sets of threads running in parallel, e.g. FFT, MAC $+$ beamformer and network transfer are on 
   different threading blocks.  The number of threads allocated to each task is adjusted to achieve optimal 
   load balancing.

Further, in order to get optimal compute performance in tasks which are highly I/O intensive, the application
needs to be tuned to (i) fulfill a majority of data accesses from processor cache (ii) reduce memory latency
to obtain peak memory bandwidth.  For this we have taken the following steps, illustrated by the section of
the code that does data reordering, delay correction, followed by FFT and fringe/FSTC phase correction.
First, data loading into cache is performed in contiguous blocks of size equal to a cache line (64 bytes).
Further, we have taken care of proper data alignment to prevent data split across the cache boundary, as
data unaligned to cache line boundary leads to double memory access.  For the code that fetches the 4
antennae data, instead of fetching the samples one at a time, we load them in chunks that are integer
multiples of 4 bytes for the data reordering loop to operate on, which eliminates 3 extra cycles of memory
access.  The reordered data for the full buffer are loaded back into memory.  When reading this data back
for delay correction and FFT, it is read in for one antenna at a time, in units that are multiples of
the FFT lengths.  This ensures optimal cache locking, as the successive operations of conversion of
8 bit samples to 16 bits, 16-bit fixed point FFT, fringe and FSTC correction using pre-loaded phase
look-up table are all implemented for these smaller segments of data.  Further, the phase corrected
spectral data are directly loaded into memory using non-temporal store instructions, which in turn helps
to reduce the cache pollution.  As a result of all these optimizations, we have reduced the total memory
I/O time to an extent that it takes only about 66\% of the data buffer time, assuming the specified
rate of $\sim$ 5 GB/s for the processor to memory bandwidth.

\subsection{Vectorized correlator}
\label{sec:12}
Performance is a function of processor clock frequency and number of instructions executed per clock 
cycle. For a given processor architecture, performance of a code can be increased by reducing the number 
of instructions it takes to execute specific tasks. The SIMD technique allows for code vectorization 
via the use of a single instruction stream capable of operating on multiple data elements in parallel. 
For example, the 128-bit streaming SIMD extension (SSE) instructions supported by Intel's Xeon processors,
enable simultaneous processing of 16 ADC data samples, each of 8 bits width.  This improves the efficiency 
of data reordering section of the GSB code. For in-place FFT and fringe correction, where the data samples 
are 16-bit wide for the fixed point version of the GSB code, 8 data elements can be processed simultaneously. 
There are 4 multiplications executed in a given instruction cycle. For this to happen successfully, the 
manipulation of data needs to be done with vector instructions, aligned at the 128-bit boundaries. 
Further, special SSE instructions are  available that flush the phase corrected spectral data from the 
processor's registers directly to main memory, without going through the cache, thus minimizing cache pollution.
For the MAC operations, the input data buffer is arranged so as to optimize cache blocking. The full
vectorized MAC loads 4 frequency channels (complex samples) for a given antenna in a single SSE data
register. This allows 4 MAC operations (complex multiplication followed by complex addition with the sum 
being in register) to be carried out simultaneously, handling 4 frequency channels. After few integrations 
it is advantageous to write back the summed product directly into the memory to maintain the cache coherency 
and load the next set of 4 frequency channels which are still in the cache.

\begin{figure}
\begin{center}
  \includegraphics[angle=0,width=0.6\textwidth]{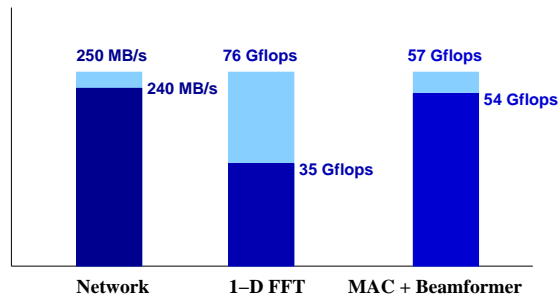}
\caption{Bar diagram to illustrate the optimized performance of the GSB code, for 3 different aspects.  The
values on top of the bars are the theoretical peak values for a single compute node, assuming dual gigabit
connectivity for the network performance and integerized vector processing for the FFT and MAC $+$ beamformer
performance. The values listed on the side of the bars are the actual, achieved figures.}
\label{fig:6}
\end{center}
\end{figure}


To reduce the memory bandwidth requirements and gain maximum compute benefits, the GSB real-time code is
tuned to operate in the fixed point integer domain. The theoretical compute power available for integerized
vector processing on a single layer-2 compute node is 76 Gflops for FFT (which uses 4 threads) and 57 Gflops
for MAC $+$ beamformer (which uses 3 threads). Our optimized {\it real-time} code is benchmarked to give a
sustained performance of 35 Gflops (i.e. $\sim$ 46$\%$ of theoretical peak) for FFT and 54 Gflops
(i.e. $\sim$ 94$\%$ of theoretical peak) for MAC $+$ beamformer, on a single compute node
(see Fig. \ref{fig:6}). These are purely compute benchmarks, and do not include any I/O overheads.
The factor of two reduction in the performance of the FFT is due to the fact that the fixed point IPP
based FFT actually uses floating point arithmetic for its internal stages.  The total achieved {\it vector}
compute power for the full 16 nodes of layer-2 is 1.4 Tflops.  This achieved compute power is almost 3
times of the real-time requirement of 490 Gflops.  This is because, as explained in Sect. \ref{sec:11}, 
about 66\% of the time goes in I/O operations and the pure compute time available is only 33\%.
This signifies that without code vectorization, we would have been more than factor of 2 off from the
real-time requirements.

\section{Performance validation of the GSB}
In the following we describe some of the main steps which have been carried out to test and validate 
the performance of the GSB, to get it ready for release as a regular backend for the GMRT.

\subsection{Numerical precision of the GSB code}
\label{sec:13}

As explained earlier, the GSB code supports both 32-bit floating point and 16-bit fixed point operation.
The primary trade-offs between these modes are that the floating point has better precision and higher
dynamic range than the fixed point, and also has a shorter development cycle, since one doesn't generally
need to worry about issues such as overflow, underflow, and truncation/round-off errors, which
can reduce the accuracy of the fixed point code.  On the other hand, fixed point code is almost always
computationally faster, especially in the case of vectorized processors.  Hence,  to optimize 
performance, the real-time version of the GSB code primarily uses fixed point computations
upto the MAC stage, after which the data are converted to floating point for the long-term accumulation.
The off-line version of the GSB code is a fully floating point processing pipeline.
In the real-time code, the signal level is tuned to minimize the effect of finite word length, i.e the
overflow on the most significant bits or quantization on the least significant bits.
Similarly, the integer fringe look-up table is properly scaled to fit the phase corrected spectral data
into the lower significant half of a 32-bit register, after 16-bit multiplication.  This eliminates the
possibility of introducing bias by truncation.  The correlation product is 32-bit wide, which is finally
accumulated in floating point form, to avoid the chances of overflow during longer integrations.

\begin{figure}
\begin{center}
\includegraphics[angle=270,width=0.8\textwidth]{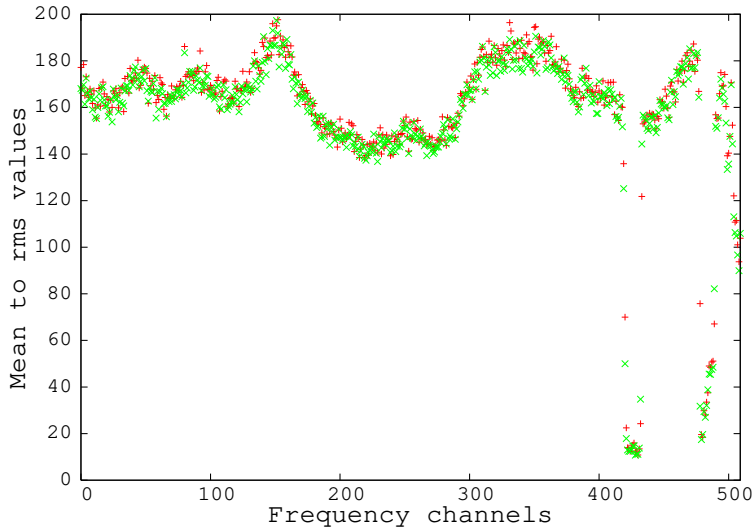}
\caption{Mean to rms ratio as a function of channel number for the auto spectrum from a single antenna, 
using data from a point source calibrator, 3C147 : green ``$\times$'' symbols are for fixed point code, 
red ``$+$'' symbols are for floating point code.  The results are from data of 18 minutes duration, 
where the individual data samples have time and frequency resolutions of 2 seconds and 32.55 KHz.}
\label{fig:7}
\end{center}
\end{figure}

\begin{figure}
\begin{center}
\includegraphics[angle=270,width=0.8\textwidth]{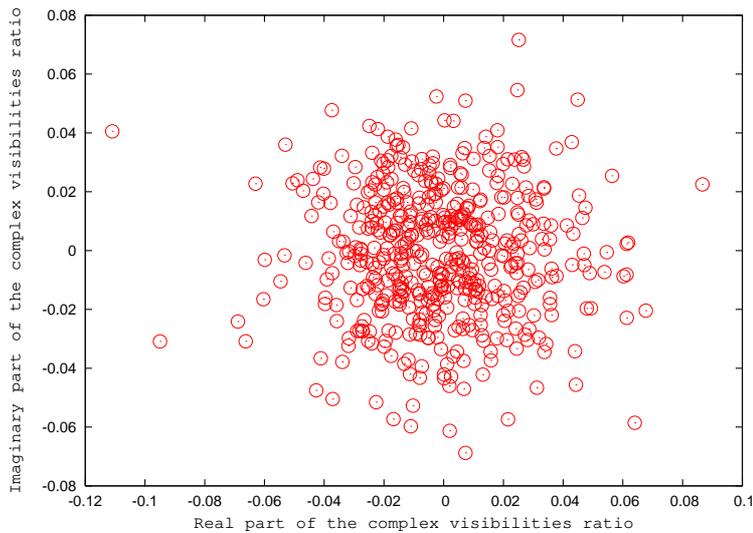}
\caption{Ratio of complex visibility values from floating and fixed point code, after subtraction a dc
of 1.0 from the real part of the ratio.}
\label{fig:8}
\end{center}
\end{figure}


To check the numerical accuracy of the fixed point code, we have compared its results with those
from the floating point code, using a recorded sample baseband data sets.  Fig. \ref{fig:7} shows 
an example, where the ratio of mean to root-mean-square (rms) power as a function of spectral channel 
number for the auto spectrum of a single antenna, using the point source calibrator quasar 3C147 as 
a test source, is used as a comparison diagnostic. The individual data samples are of 2 seconds time 
resolution and 32.55 KHz frequency resolution, and the total data span used for the computations is 
18 m. The results from the fixed point and floating point codes are quite similar (including the 
degradation due to RFI near channels 420 and 480), except for a small degradation ($\sim$ 1.5\%) of 
the mean to rms ratio for the fixed point results compared to the floating point, which is due to a 
small increase in the rms due to the fixed point arithmetic.

Fig. \ref{fig:8} shows a more direct and detailed comparison where the ratio of the complex visibility 
values obtained for the fixed and floating code is used as a diagnostic. The center of the distribution 
is shifted from [1,0] to [0,0] by removing the dc from the real part of the ratio.  In the ideal case of
no difference between the fixed and floating point calculations, all the data points would have ended up 
at [0,0] in this plot.  The observed distribution shows a fairly symmetric spread around [0,0], with a rms 
of 0.02. There is a small offset in the mean value of less than 10\% of the rms, which is due to slight
inaccuracies in the scaling factors for the fixed and floating point pipelines. However, this is not of 
relevance here, as the numerical accuracy of the fixed point to the floating point comparison is 
represented by the spread of the numbers after division, rather than by the offset. This 
result again signifies a  degradation (of $\sim$ 2\%) in going from floating point to fixed point 
arithmetic.  This small degradation is a minor price for the significant computational speed-up 
that the fixed point code provides.

\subsection{Validation of the GSB in correlator mode}
\label{sec:14}
In order to validate the correlator mode of the GSB, we have compared the results from our off-line
code with those from the Swinburne DiFX correlator \cite{Deller} running on the same baseband recorded data 
sets. For validating our real-time code, we have carried out extensive test runs with the GMRT, having
the hardware backend running in parallel with the GSB.  Some of the results from these tests and 
comparisons are presented below.

\subsubsection{Validation of the off-line version of the GSB}
\label{sec:15}

\begin{figure}
\begin{center}       
\includegraphics[angle=270,width=0.8\textwidth]{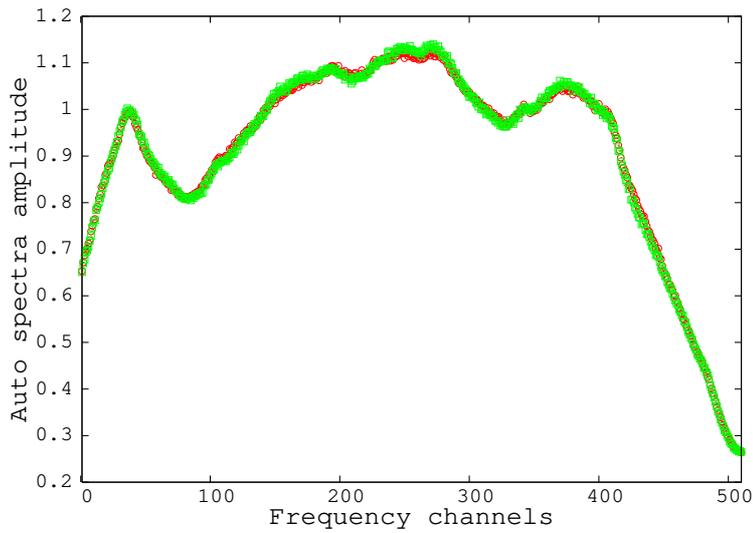}
\caption{Auto spectra for a single GMRT antennae for the data on calibrator 3C147 at 610 MHz, 
as a function of channel number : green, open boxes are for the GSB; red, open circles are for the DiFX. 
Each spectral point is of 2 seconds time and 32.55 KHz frequency resolution.}
\label{fig:9}       
\end{center}       
\end{figure}

\begin{figure}
\begin{center}       
\includegraphics[angle=270,width=0.8\textwidth]{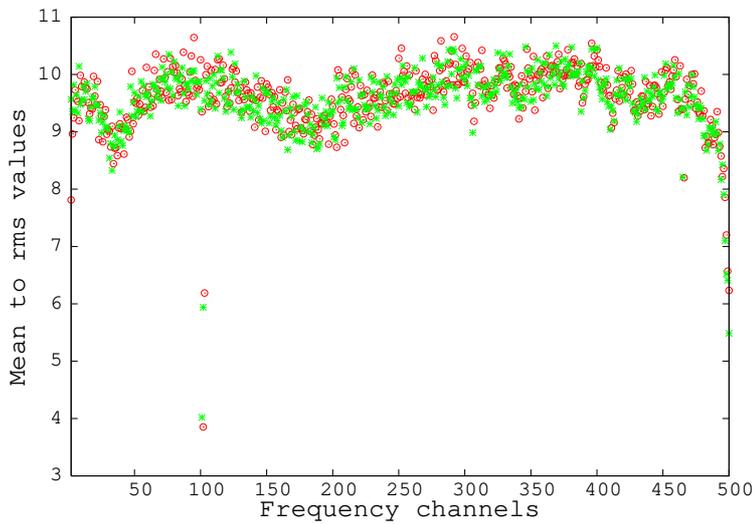}
\caption{Mean to rms ratio for different spectral channels of the cross spectrum for a single GMRT baseline 
for the data on calibrator 3C147 at 610 MHz : green, stars are for the GSB; red, open circles are for 
the DiFX.  The individual data samples are of 2 seconds time and 32.55 KHz frequency resolution.  The mean and rms
are computed from 13 minutes of data.}
\label{fig:10}      
\end{center}       
\end{figure}

GSB baseband data recorded for the point source calibrator 3C147 was used to compare the 
performance between the GSB (off-line floating point code) and the DiFX, both running on the Itanium 
cluster at the NCRA.  Self and cross visibility spectra were computed for both codes, with time and 
frequency resolutions of 2 seconds and 32.55 kHz, for the 13 minutes duration of the test data.  
Sample auto spectra and cross spectra from individual baselines, were used as the comparison 
diagnostic (see Fig. \ref{fig:9} and \ref{fig:10} for examples of this).  Although these 
pipelines have very different implementations, with different delay correction models and different 
fringe correction procedures (DiFX uses CALC \footnote{see http://www.gemini.gsfc.nana.gov/solve} 
based delay generation and pre-FFT fringe correction), there are no significant differences seen in the 
results.  The overall shapes of the 2 seconds integrated auto spectra, after scaling by a constant factor 
to match the amplitudes, agree very well (Fig. \ref{fig:9}).  The rms of the intensity fluctuations in 
individual spectral channels of these spectra, calculated over the 13 minutes duration of the test data, were
also found to match to within 1\% level.  Fig. \ref{fig:10} shows a comparison of the signal to noise ratio 
(mean visibility amplitude divided by the rms deviation) for different spectral channels, for a given 
baseline from the data set.  The mean and rms were computed from the 13 minutes duration of data.  The
agreement is found to be very good. All these comparisons provide an independent and valuable check of 
the GSB computational pipeline.

\subsubsection{Tests of the real-time version of the GSB}
\label{sec:16}

\begin{figure}
\begin{center}
  \includegraphics[angle=270,width=0.8\textwidth]{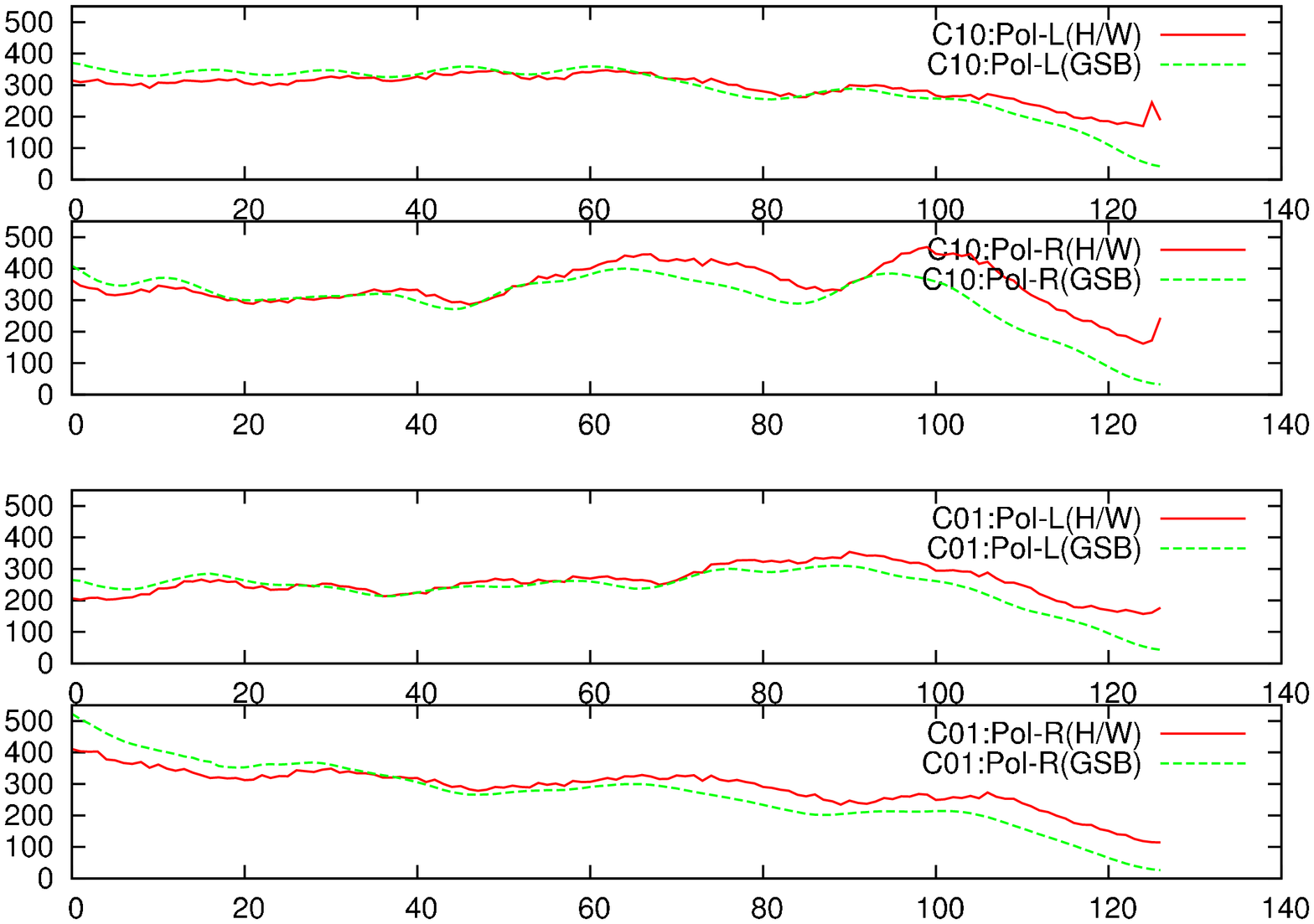}
\caption{Sample auto spectra covering 1280-1296 MHz on the calibrator source J1609+266 : data recorded
simultaneously with the GMRT hardware correlator with 16 MHz bandwidth, 128 channels and the GSB in real-time
mode with 16.66 MHz bandwidth, 256 channels.  GSB spectra have been integrated over two adjacent channels to
allow for better comparison with hardware correlator data.}
\label{fig:11}
\end{center}
\end{figure}

We have carried out extensive tests to validate the performance of the real-time code of the GSB,
including comparisons with the hardware backend running in parallel. Here we report a sample comparative
study that was done using observations at L-band (1280-1296 MHz) of the GMRT for a duration of $\sim$ 
8.5 hrs on a quasar J1609+266. The GSB was operated in 16 MHz bandwidth mode with 256 spectral channels 
across the band, whereas the GMRT hardware correlator generated 128 spectral channels across the same band.  
Both the correlators were configured to compute 16 seconds integrated visibility products for both polarizations.  
Fig. \ref{fig:11} compares the auto spectra produced by these two backends, for 2 GMRT antennae.  The pass 
band shapes are very similar. However, spectra from the hardware correlator show small scale undulations, 
whereas the GSB spectra are much smoother in nature. This illustrates the improved performance of the GSB 
with respect to quantization effects due to limited precision that affect the hardware backend -- the GSB
with 16-bit fringe multiplier has comparatively less quantization noise. The typical effect of this is 
$\sim$ 20 to 50 \% improvement in the rms noise in the final images made with GSB data. Further detailed 
comparisons between the GMRT hardware backend and the GSB, to refine the quantitative measure of the improvement, 
are expected to be carried out once the GSB is available for regular use.  

\begin{figure}
\begin{center}
  \includegraphics[angle=270,width=0.8\textwidth]{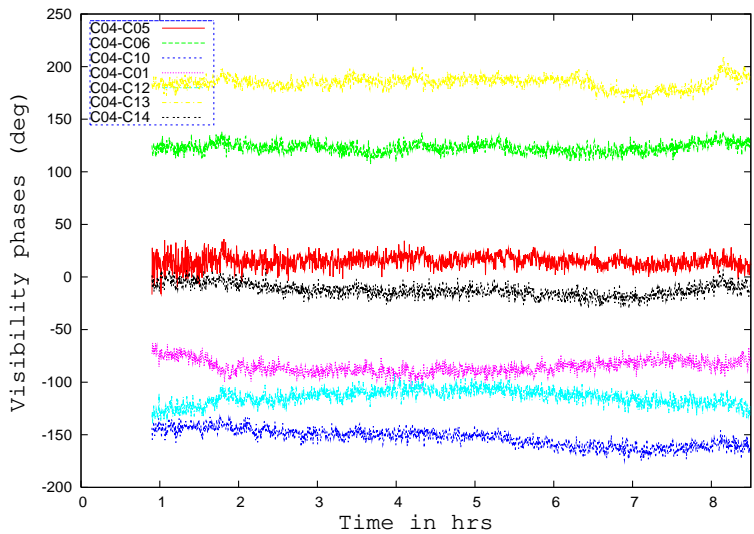}
\caption{Visibility phases for a single spectral channel, for a sample of selected baselines, for a
duration of 8.5 hrs : data recorded with the real-time mode of the GSB, on the calibrator source J1609+266.}
\label{fig:12}
\end{center}
\end{figure}

\begin{figure}
\begin{center}
  \includegraphics[angle=0,width=1.0\textwidth]{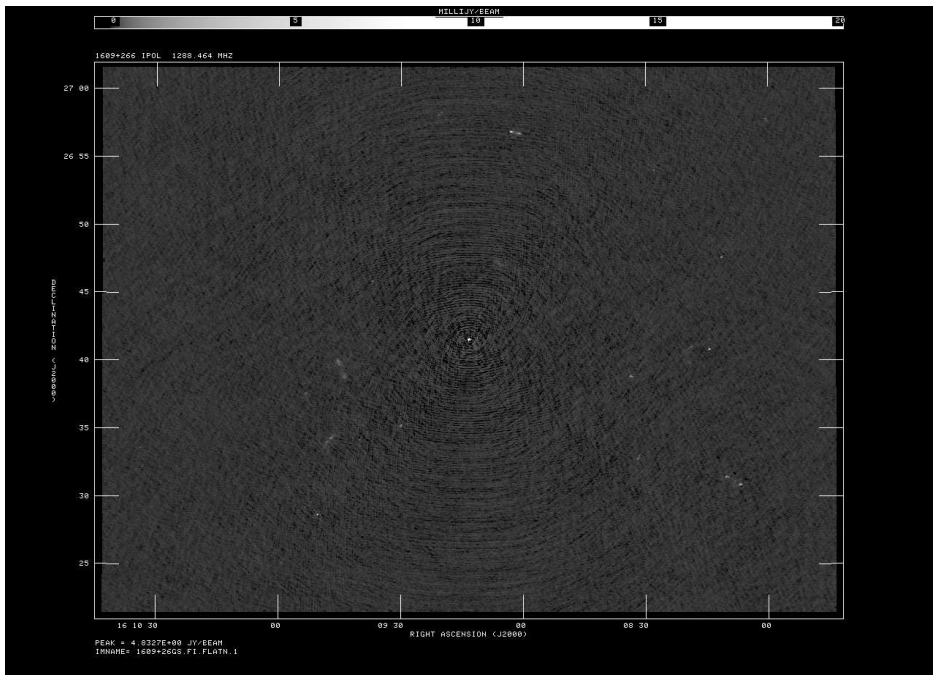}
\caption{A first light image from the GSB : radio intensity map of the region of the sky around the source
1609+266, at 1280 MHz. Beside the central point source J1609+266 (4.8 Jy), a few other sources, including
a couple of double radio sources (white dots) can be clearly seen.  Dynamic range achieved in this image
is $1.4\times10^{5}$. (Courtesy : Subhashis Roy)}
\label{fig:13}
\end{center}
\end{figure}

The long-term stability of the results from the GSB are illustrated in Fig. \ref{fig:12}, which shows the 
visibility phases from the GSB data for 7 baselines over a duration of 8.5 hrs. The phases show very 
small variations ($\sim$ $\pm$ 10 deg) over the entire duration.  The final image of the field made using 
the full duration GSB data is shown in Fig. \ref{fig:13}.  The central strong source, J1609+266, has a
strength of 4.83 Jy.  Signal detected at half power beam width (15') away is 34 $\mu$Jy, which is 1.5 times 
the thermal limit.  The dynamic range achieved in this map is 1.4$\times 10^{5}$. 

\begin{figure}
\begin{center}
  \includegraphics[angle=270,width=0.8\textwidth]{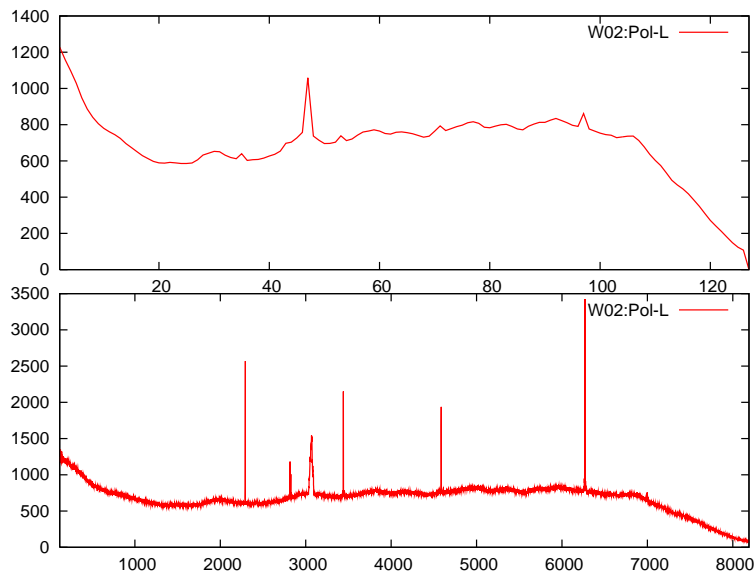}
\caption{Illustration of the high spectral resolution capability of the GSB : sample auto spectra on the
source 3C48, at 244 MHz, with 128 spectral channels (top panel) and 8192 spectral channels (bottom panel).
Both the spectra are of 2 seconds time resolution.}
\label{fig:14}
\end{center}
\end{figure}

\begin{figure}
\begin{center}
  \includegraphics[angle=270,width=0.8\textwidth]{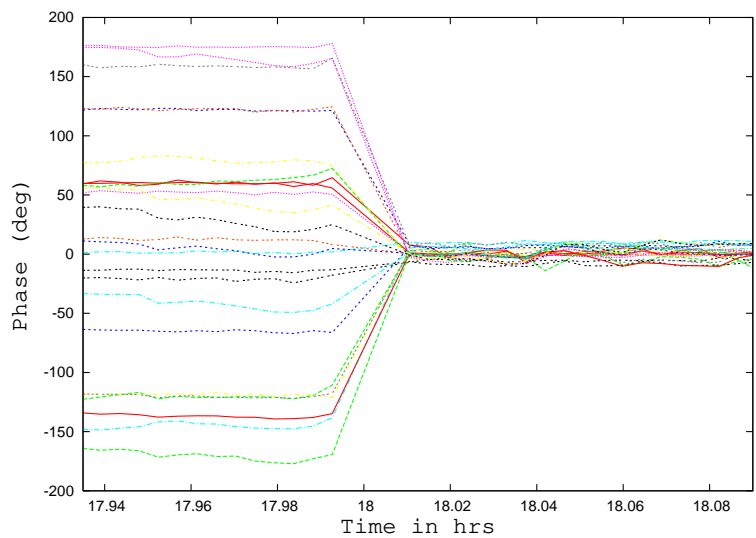}
\caption{Demonstrating the real-time phasing of the array with GSB data : antenna based phase offsets
for a single spectral channel, for several antennae, before phasing (on the left) and after phasing
(on the right).}
\label{fig:15}
\end{center}
\end{figure}

In Fig. \ref{fig:14} we show an example of the enhanced spectral resolution capabilities of the GSB.
Full bandwidth, high resolution spectra with 8192 spectral channels from the GSB are compared against the
regular 128 channel spectra.  The data were taken on the calibrator source 3C48 at 244 MHz. The better
detectability of the narrow spectral features in the high spectral resolution data is clearly
demonstrated, allowing possibilities for better excision of such unwanted signals.
            
\subsection{Results from the GSB beamformer}
\label{sec:17}

As discussed in Sect. \ref{sec:3}, the GSB beamformer is designed to produce incoherent and coherent
array beams simultaneously.  The beam formation involves addition of pre or post detected samples from 
individual antennae. Prior to addition, the power levels at the sampler inputs need to be equalized, 
which can be done with the help of programmable attenuators in the GMRT analog receiver chain. In addition, 
the GSB has its own variable gain amplifiers at the input of each sampler which are used for finer 
adjustments. Further, for coherent beam formation, we need to calibrate out the antenna based phase offsets 
before the voltages can be added.  Antenna based phases are solved for using the recorded cross-correlations 
on a calibrator source, and then these phases are applied after the FFT stage, as an additional term in
the fringe and FSTC corrections. Fig. \ref{fig:15} demonstrates the phasing of the array.

\begin{figure}
\begin{center}
  \includegraphics[angle=270,width=0.8\textwidth]{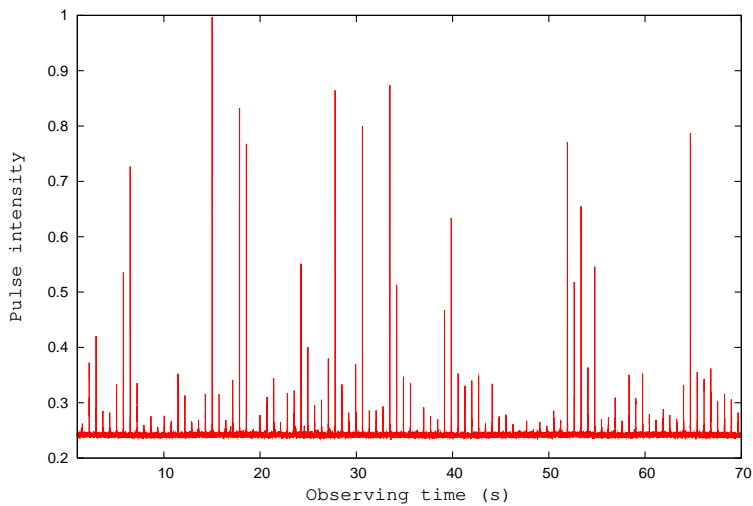}
\caption{Single pulse time series for PSR B0329+54 at 610 MHz, illustrating the incoherent array beam 
forming mode of the GSB.  Intensity signals from both polarizations of 15 antennae were added.  The pulsar
has a period of 714.5 ms, there are around 97 single pulses seen over this 70 seconds duration.}
\label{fig:16}
\end{center}
\end{figure}

\begin{figure}
\begin{center}
  \includegraphics[angle=270,width=1.0\textwidth]{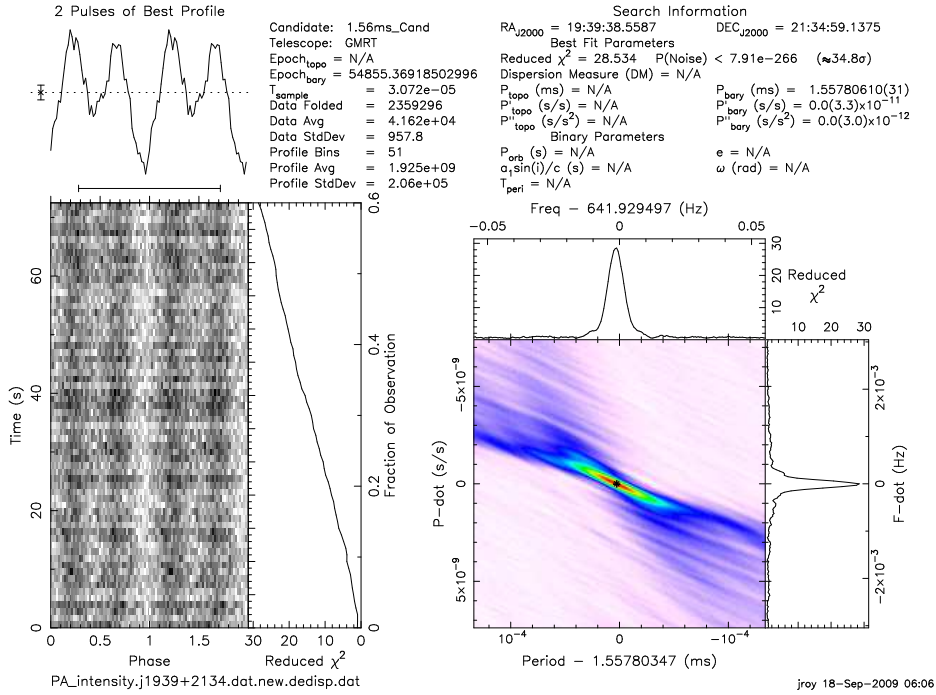}
\caption{Dedispersed and synchronously folded profile of PSR B1937+21 at 325 MHz, from phased array GSB
data, using the PRESTO analysis package.  Pulsar main pulse and inter-pulse seen in the gray scale plot
are significantly smeared due to residual dispersion within single spectral channel. The data is plotted
over 2 pulse phases.}
\label{fig:17}
\end{center} 
\end{figure}

\begin{figure}
\begin{center}
  \includegraphics[angle=270,width=1.0\textwidth]{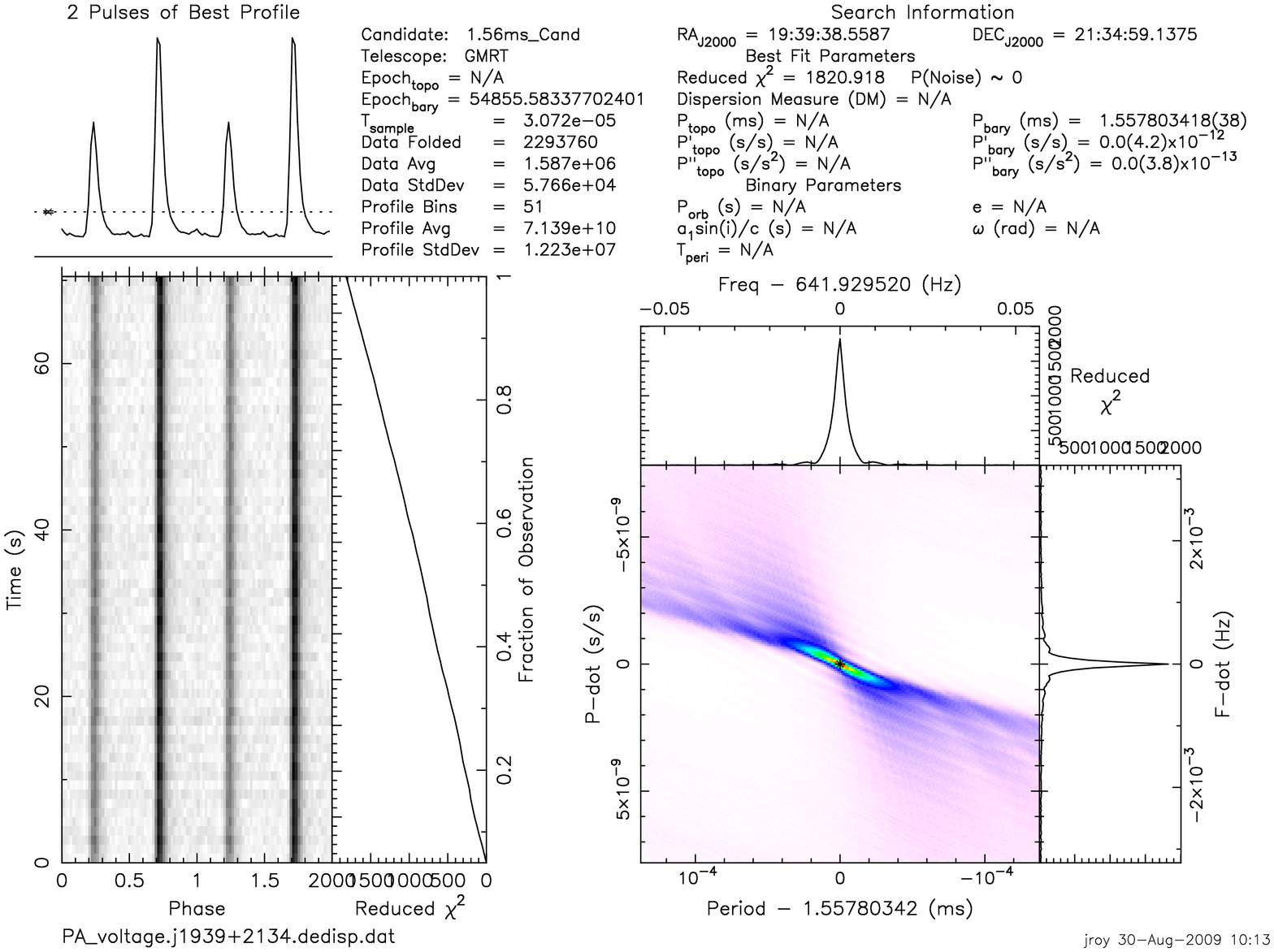}
\caption{Coherently dedispersed profile of PSR B1937+21 at 325 MHz, with a time resolution of 30 $\mu$s,
folded using PRESTO. Pulsar main pulse and inter-pulse are now seen clearly as separate, sharp pulses.
There is a small indication of scattering tail.}
\label{fig:18} 
\end{center}
\end{figure}

As an example of the incoherent array mode of operation of the GSB beamformer, Fig. \ref{fig:16} shows
the single pulse time series for the pulsar B0329+54 at 610 MHz.  Signals from both polarizations of
15 antennae were added after conversion to intensity in the beamformer. There were 512 channels across
the bandwidth of 16.66 MHz, and the sampling rate was 30 $\mu$s (which is a factor of 4 better than the
best time resolution achievable with the hardware backend).  During off-line analysis, the recorded data
were incoherently dedispersed to remove the effect of interstellar dispersion using the
PRESTO \footnote{http://www.cv.nrao.edu/~sransom/presto/} analysis package. The resulting time series
clearly shows strong, single pulses.  This demonstrates the basic operation of the GSB beamformer.

Fig. \ref{fig:17} shows a sample result from the coherent phased array mode of operation of the GSB
beamformer.  PSR B1937+21, one of the fastest known millisecond pulsars with a periodicity of 1.55 ms,
was observed at 325 MHz using 16 antennae, after phasing the array on a calibrator source.  The voltage
signals for each polarization were summed separately in the GSB, and the resulting signals were converted
to intensities and added to obtain the total power signal.
The data were recorded with 256 spectral channels across 16.66 MHz (65.1 KHz spectral resolution),
with 30.72 $\mu$s time resolution.  During off-line analysis using PRESTO, the data were incoherently
dedispersed and then folded synchronously with the Doppler corrected topocentric pulsar period to
obtain the average pulse profile.  The top left panel of the figure shows this average profile as
2 consecutive pulses.  The profile has 51 time bins, and shows two emission components which appear
quite broad and overlapping with each other because of the uncorrected dispersive effects within the
individual spectral channels.  For this pulsar, with a DM of 71.0398 pc/cc, this residual smearing
at 325 MHz for 65.1 kHz spectral resolution works out to be 1.1 ms -- this is substantial, compared
to the pulsar period.

The GSB has the capability to improve upon the above situation by using coherent dedispersion, as it
provides for a mode where the pre-detected voltages from the coherent array sum for each polarization
can be recorded for off-line processing.  Our coherent dedispersion pipeline first converts these
spectral voltage data into the corresponding broadband time-series, by carrying out an inverse FFT
operation.  The resulting voltage time series data are then put through the deconvolution process where
we correct for the dispersion by multiplying the input data with the inverse transform of the ISM
dispersion kernel \cite{Hankins} in the Fourier domain.  The corrected voltage time series data for
both polarizations are then converted to intensities and added together to produce the total power
time series, which is integrated to the desired time resolution.  The improvement achieved from this
processing is illustrated in Fig. \ref{fig:18}, which shows the coherently dedispersed profile for 
PSR B1937+21, reduced to the same final time resolution (30 $\mu$s) as in Fig. \ref{fig:17}.  
The improvement in the shape of the profile between these two figures is striking, and illustrates 
the power of coherent dedispersion, which is one of the enhanced capabilities that the GSB provides.

\begin{figure}
\begin{center}
  \includegraphics[angle=270,width=0.8\textwidth]{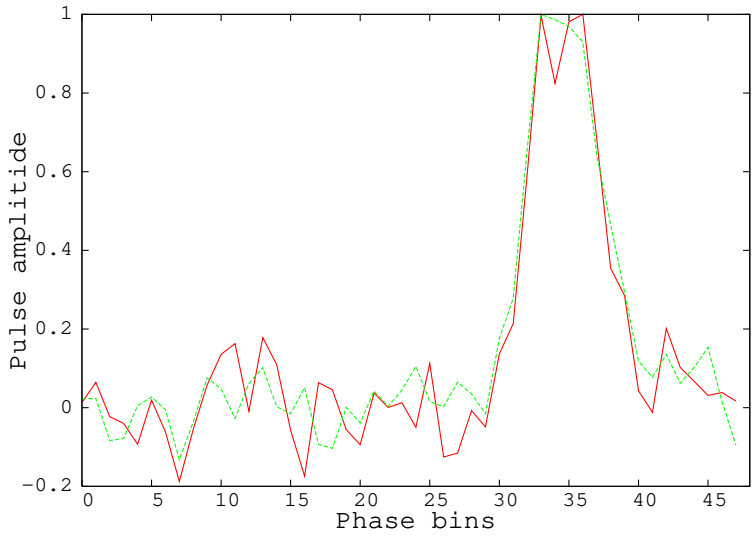}
\caption{Dedispersed and synchronously folded profiles of PSR J1300+1240 at 610 MHz, from phased array
observations with 20 GMRT dishes.  The profile from the GSB data is the green, "dotted" line, whereas
the profile from the hardware backend data is the red, "solid" line. The time and frequency resolutions,
as well as the total length of data used, have been matched for both the backends.  The off-pulse mean
values were subtracted from the original profiles and the these were scaled to the same amplitude to
make the comparison.}
\label{fig:19}
\end{center}
\end{figure}

For a quantitative comparison between the hardware backend and the GSB, Fig. \ref{fig:19} shows the
pulse profiles obtained for a millisecond pulsar J1300+1240, whose period is 6.22 ms. The same set of
20 GMRT antennae were added in both the hardware backend and the GSB phased array beamformer, and the
data recording was exactly concurrent in time. The hardware backend data have 256 spectral channels
(62.5 KHz spectral resolution) and 128 $\mu$s time resolution, while the GSB data have 512 spectral
channels (32.55 KHz spectral resolution) and 30.72 $\mu$s time resolution. In order to compare with the
hardware backend, the GSB data were degraded by a factor of 2 in spectral resolution and a factor of 4
in time resolution.  The final folded profiles shown in Fig. \ref{fig:19} are quite similar. The off-pulse 
mean values were subtracted from the original profiles and the these were scaled to the same amplitude to
make the comparison. While the on-pulse deflections are same for the original profiles, there is a $\sim$ 30\% 
lower off-pulse rms estimated for the GSB data with respect to the hardware backend. This is similar to the 
typical improvement in rms achieved for the imaging data with the GSB.

\section{Enhanced capabilities of the GSB}
\label{sec:18}

Besides providing enhanced flexibility in the basic parameter space, such as better time and
frequency resolution and coherent dedispersion capabilities (which have been demonstrated in
the previous sections), the GSB also holds the promise of adding several new capabilities to
the GMRT data processing pipeline.  One of the most useful of these is the ability to add new
algorithms for RFI detection and mitigation.  RFI is an ever increasing problem for radio astronomy,
especially at the lower frequencies which the GMRT operates at.

There are several approaches to RFI detection and mitigation.  As a preliminary attempt, aimed to
tackle the impulsive, broad-band RFI that is more often seen at the GMRT, we have implemented a
time-domain RFI blanker that acts on the pre-detected voltage data.  We use the Median of Absolute
Deviation (MAD) to derive an estimator for discriminating the outliers in the voltage data stream,
rather than the more traditional approach that uses the variance.  This is because the MAD estimator
is more robust in the presence of large outliers such as would be produced by strong RFI spikes. The
MAD for a block of N data points is defined as :
\begin{equation}
MAD = median { |x_{i} - \tilde{x}| }
\end{equation}

where $\tilde{x}$ is the median for the N data points represented by the voltage samples, $x_{i}$.
The MAD thus estimates the median of the absolute value of the deviation of individual samples from
$\tilde{x}$.  The threshold for discrimination is derived from the MAD estimator, assuming the
underlying distribution for the RFI free voltage samples to be a normal distribution, and can be
expressed as
\begin{equation}
|x_{i} - \tilde{x}| > M \frac{MAD}{0.6745}
\end{equation}

where $M$ is a user settable parameter, with typical values of 2.5 to 3.0. The data samples crossing
the threshold are replaced by pre-computed Gaussian distributed noise samples, which are ensured to
be independent for different antennae.

\begin{figure}
\begin{center}
  \includegraphics[angle=270,width=0.8\textwidth]{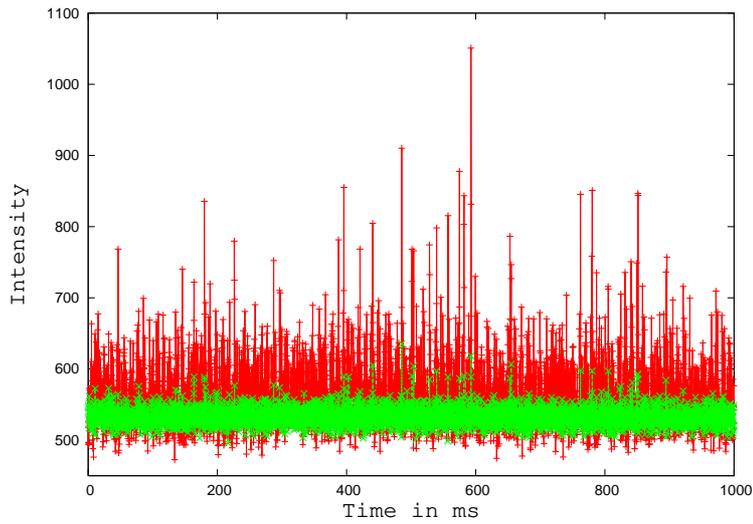}
\caption{Removal of impulsive RFI using MAD technique : Total intensity of a single antenna on calibrator
3C48 at 156MHz, with 1 ms time resolution : green, ``$\times$'' symbols represent data after filtering;
red, ``$+$'' are data points before filtering.}
\label{fig:20}
\end{center}
\end{figure}

\begin{figure}
\begin{center}
  \includegraphics[angle=270,width=0.8\textwidth]{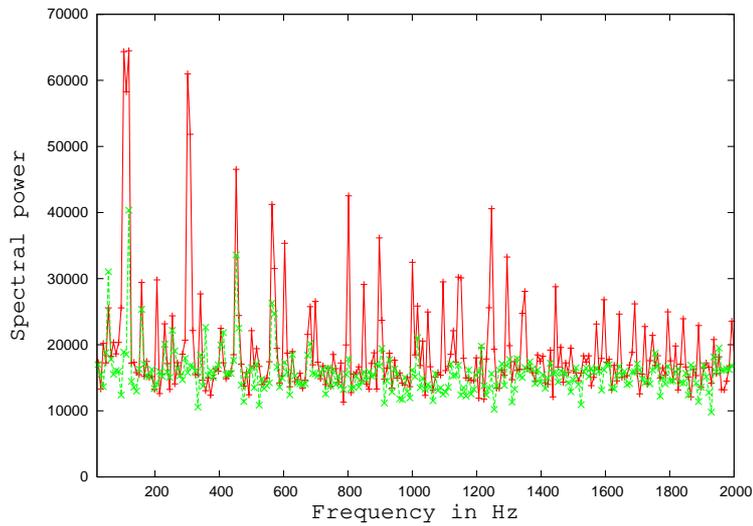}
\caption{Removal of impulsive, powerline related RFI using MAD technique : Spectral content of the intensity
signal from a single antenna on calibrator 3C48 at 156MHz : green, ``$\times$'' symbols represent data after
filtering; red, ``$+$'' are data points before filtering.}
\label{fig:21}
\end{center}
\end{figure}

\begin{figure}
\begin{center}
  \includegraphics[angle=0,width=0.8\textwidth]{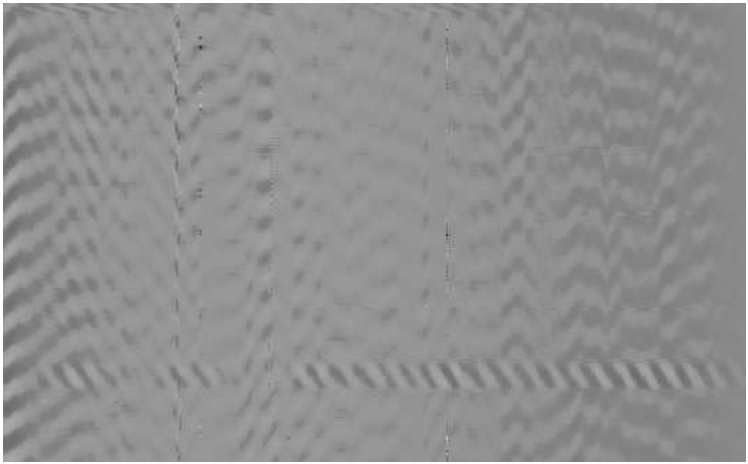}
\caption{Gray scale display of cross-visibility amplitude for a short GMRT baseline as a function of time
(y-axis) increasing downwards and frequency (x-axis) increasing to the right, on a calibrator 3C48 at 156MHz.
This is before applying the RFI filtering. The residual fringe patterns due to terrestrial RFI signals are visible.}
\label{fig:22}
\end{center}
\end{figure}

\begin{figure}
\begin{center}
  \includegraphics[angle=0,width=0.8\textwidth]{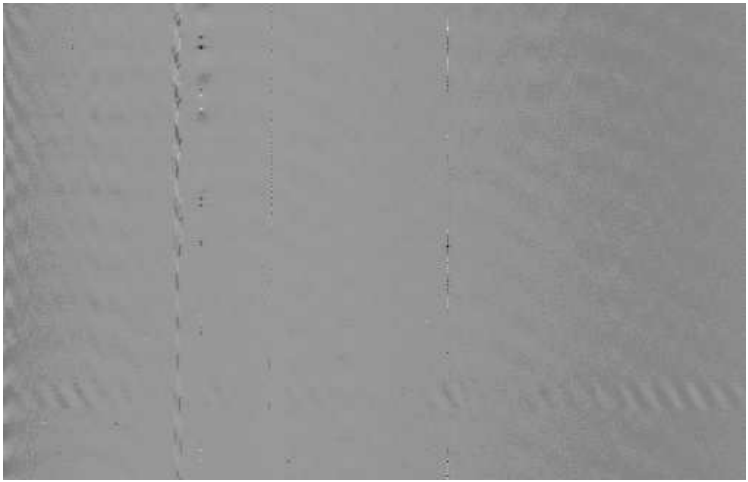}
\caption{Same as Fig. \ref{fig:21}, showing the improvement after applying time-domain MAD filtering}
\label{fig:23}
\end{center}
\end{figure}

Fig. \ref{fig:20} shows an example of testing this algorithm.  GMRT data recorded at 156 MHz was chosen,
as it is usually very much affected by RFI.  This can be seen in the figure as the large spikes in the
total intensity data, which is from a single antenna and has been averaged to 1 ms time resolution
(red curve with ``$+$'' symbols).  Most of these RFI spikes are attributable to spark discharges from
power lines, which are broadband and quasi-periodic in nature.  This can be seen in the power spectrum
of the intensity signal (Fig. \ref{fig:21}), which exhibits significant power at 50 Hz, 100 Hz and higher
harmonics. The results from the MAD filtering of the raw voltage data are also shown in Fig. \ref{fig:20}
and Fig. \ref{fig:21} (green curves with ``$\times$'' symbols) $-$ the improvement in signal quality is
quite significant.  The fraction of data samples replaced by the filtering is only 2-3\%.

Fig. \ref{fig:22} presents the amplitude of fringe and delay corrected cross-visibility data for a short
GMRT baseline, plotted in frequency-time space, from the same 156 MHz observations.  All the patterns
with periodic modulations seen in this figure are due to underlying RFI.  The modulations are due to the
fact that terrestrial sources get fringe rotated at the fringe frequency applied for the sky signal.
The MAD based RFI excision is also effective in filtering out most of these features (Fig. \ref{fig:23}),
leaving behind a much cleaner data set.  The raw data are off from the thermal-limit by factor of 10,
whereas the filtered results (both the self power and cross-visibility data) are off only by a factor
of 2.5.  This results in an improvement in sensitivity by a factor of 4, with a very small amount of data
being discarded.  There is significant scope for further improvements in filtering techniques that can
be deployed with the GSB.

Amongst other enhanced capabilities that the GSB can provide, the gated correlator mode is of
significant interest and potential.  Besides allowing the possibility of detecting low-level, off-pulse
radio emission from the pulsars, this mode can allow pulsars, which are ideal point source phase
calibrators, to be used for calibrating visibilities (including in-field calibration capabilities).
This has been recently demonstrated for the GSB, as part of the GMRT EoR (Epoch of Reionization)
experiment \cite{pen}.  In a similar vein, the GSB software can easily be adapted to provide the
additional facility of gain calibration using a switching noise source at the front-end of each antenna.
The GSB can also be used as a RFI localization tool, employing specialized near field imaging techniques
to map out the location of RFI sources around the observatory.  Another specialized mode of the GSB that
is expected to have wide ranging applications is the raw voltage recording mode.  Furthermore, the capability
of producing multiple beams $-$ either from different sub-sets of antennae in incoherent addition mode, or
multiple phased array beams covering different directions within the primary beam $-$ has significant
potential for wide-field searches for pulsars and transients.

\section{Future prospects}
\label{sec:19}

The GSB, even with its enhanced features and extended parameter space, had a development cycle of about
3.5 years, whereas hardware correlators like the GMRT hardware correlator or the EVLA and ALMA correlators,
typically have much longer design and development cycles, extending up to 10 years. Considering that
both hardware and software correlators benefit from Moore's law, the 10 year design phase spans a factor
of 128 for the hardware system, whereas the software design loses by factor of 4 only.  Thus, software
backend design cycles stand to gain more from Moore's law. Hence, though the general-purpose computer
is less efficient in terms of the gate count, with the aid of highly optimized code, faster development
cycle and easy upgrade, software based processing appears to be a more cost effective solution.  Our
current cost estimate for the GSB (given in Table \ref{tab:1}) translates to \$190 per correlator baseline 
(this includes the cost of the acquisition also). Furthermore, the compute to power ratio for our GSB 
implementation works to a fairly efficient value of 260 Mflops/Watt, based on the measured power consumption 
of the layer-2 nodes and the peak compute rating achieved by them. Using the same rating, we get a 
compute-per-cost ratio of 45 Mflops/\$.

The GSB concept can be expanded to bigger sized backends.  At the GMRT, we are exploring at possibilities
of a 400 MHz bandwidth, 32 station correlator and beamformer.  The 16 ADC boards operating at 33 MHz bandwidth
for 32 antennae can easily be reconfigured to support a 4-bit, 100 MHz, 16 antennae or a 4-bit, 200 MHz, 8
antennae backend.  Based on the design experience with the 33 MHz GSB, our extrapolated figures for a 400 MHz
version (GSB-HBW) are also given in Table \ref{tab:1}. The cost of the compute engines for the current 33 MHz 
design and also for the projected 400 MHz design are tabulated in the last column.  The cost for the 400 MHz 
design is for 2011 estimates, considering a factor of 4 performance gain due to operation of Moore's law.
With present technology, the 400 MHz backend will require ten-fold more resources than our current GSB system.
In order to benefit from the Moore's law predicted cost/performance gain, we also need to
adapt new computing technology.  Looking towards the future, Intel's new i7 core architecture with integrated
memory controller to reduce memory latency and to operate at higher memory bandwidth (25 GB/s) will be a big
aid for data-intensive higher bandwidth backends. There is also expected to be a large jump in vector processing
capability with the introduction of 256-bit advanced vector extension (AVX) registers.  This will provide a 
big functional enhancement in arithmetic as well as data processing aspects. Furthermore, GPU (graphical 
processing unit) based processing is going to be a big driving force towards machines that can provide 
1 Tflop per compute node.  10G base-T ten gigabit ethernet is going to bridge the gap in high speed data sharing.  
All these upcoming advancements in the computing industry will extend Moore's law well into the next decade and, 
rather than increasing the number of processing elements linearly with bandwidth and/or number of antennae/baselines, 
they will make software signal processing using assembled low-cost clusters with GPUs more and more attractive as a 
solution to the growing needs.
 
\begin{table}
\begin{center}       
\caption{Current GSB cost budget along with the extrapolated figures for 400 MHz}
\vspace{0.3cm}
\label{tab:1}
\begin{tabular}{|c|c|c|c|c|c|c|c|c|c|c|c|c|c|c|c|c|c|}
\hline
     &Band   & \multicolumn{3}{|c|}{Beamformer} & \multicolumn{3}{|c|}{FFT+MAC} & No        & Compute cost$^\ddagger$ \\
     &Width  & Beams & Out  & Tflops$^\dagger$  & Channels & Base  & Tflops$^\dagger$  & of compute&             \\
     &(MHz)  &       & Bits &                   &          & line  &                   & nodes     &          \\ \hline
GSB  &33     &   2   & 16   & 0.017             & 4096     & 528   & 0.46              & 32        &\$56,000  \\
     &       &       &      &                   &          &       &                   &           &          \\\hline
GSB  &400    &   2   & 16   & 0.21              & 4096     & 528   & 5.8               & 200       &\$175,000 \\
HBW  &       &       &      &                   &          &       &                   &           &          \\\hline
\end{tabular}
\end{center}       
$\dagger$ : This is the figure for the required compute power.\\
$\ddagger$ : The dollar value shown is the dollars at the time of purchase.
\end{table}

\section{Summary}
\label{sec:19}

We have described our design and implementation of a 33 MHz bandwidth, 32 station, dual polarization,
fully {\it real-time} software backend system for the GMRT. Our approach has allowed relatively rapid development
of a fairly sophisticated and flexible backend receiver system, which will greatly enhance the productivity
of the GMRT.  We have successfully validated the backend and it has now been released for regular use.
We have also demonstrated some of the versatile features of the new backend, and described its capabilities
for RFI mitigation and other enhanced features.  We believe this is the first instance of a software based
${\it real-time}$ backend for an intermediate sized array like the GMRT.  Our approach holds promise for
future developments for bigger radio telescopes and wider bandwidths.

\begin{acknowledgements}

We would like to thank many of the staff members from NCRA who have helped at various stages
of implementing the GSB as an observatory backend.  We acknowledge Santaji Katore and Nilesh Raskar
for their contribution towards building a Graphical User Interface for the GSB.  We are grateful
to Subhashis Roy for help in analyzing and debugging the performance of the GSB, working on several
test data sets, and also for providing the high dynamic range image of the J1609+266 field.
We thank Sharwari Kulkarni for her contribution in implementing the narrow-band module for the GSB,
to support the high spectral resolution observing mode.  We specially thank the computer group at
the GMRT (Mangesh Umbarje and Sumit Mirajkar) for extensive help at various stages of assembling
and testing the GSB cluster and network.  We thank Rajaram Nityananda and Jayaram Chengalur for
their useful comments during the design phase.  We also thank Dipankar Bhattacharya for useful
discussions during implementation of the coherent dedispersion pipeline of the GSB. We acknowledge 
Adam Deller for helping in the comparative study of the results from the GSB with the DiFX.  
Finally, we thank B. Ajithkumar and his group at the GMRT, for building the new 32 MHz baseband system 
that feeds the analog signals to the GSB.  The GMRT is run by the National Centre for Radio Astrophysics 
of the Tata Institute of Fundamental Research.

\end{acknowledgements}

\end{document}